\documentclass[conference]{IEEEtran}
\IEEEoverridecommandlockouts
\usepackage{cite}
\usepackage{amsmath,amssymb,amsfonts}
\usepackage{graphicx}
\usepackage{textcomp}
\usepackage{xcolor}
\usepackage{comment}
\usepackage{algorithm}
\usepackage{algpseudocode}
\usepackage{booktabs}
\usepackage{placeins}
\usepackage{adjustbox}

\usepackage{url}
\usepackage{pifont}

\usepackage{subcaption}
\usepackage{multirow}

\newcommand{\inner}[2]{\langle #1, #2 \rangle}

\def\BibTeX{{\rm B\kern-.05em{\sc i\kern-.025em b}\kern-.08em
    T\kern-.1667em\lower.7ex\hbox{E}\kern-.125emX}}
\begin{document}

\title{ScalableHD: Scalable and High-Throughput Hyperdimensional Computing Inference \\ on Multi-Core CPUs}

\newcommand\blfootnote[1]{%
  \begingroup
  \renewcommand\thefootnote{}\footnote{#1}%
  \addtocounter{footnote}{-1}%
  \endgroup
}

\author{
\IEEEauthorblockN{Dhruv Parikh, Viktor Prasanna}
\IEEEauthorblockA{
   University of Southern California, Los Angeles, California, USA \\
    dhruvash@usc.edu, prasanna@usc.edu}
}

\maketitle

\begin{abstract}

Hyperdimensional Computing (HDC) is a brain-inspired computing paradigm that represents and manipulates information using high-dimensional vectors, called hypervectors (HV). Traditional HDC methods, while robust to noise and inherently parallel, rely on single-pass, non-parametric training and often suffer from low accuracy. To address this, recent approaches adopt iterative training of base and class HVs, typically accelerated on GPUs. Inference, however, remains lightweight and well-suited for real-time execution. Yet, efficient HDC inference has been studied almost exclusively on specialized hardware such as FPGAs and GPUs, with limited attention to general-purpose multi-core CPUs.
 
To address this gap, we propose ScalableHD for scalable and high-throughput HDC inference on multi-core CPUs. ScalableHD employs a two-stage pipelined execution model, where each stage is parallelized across cores and processes chunks of base and class HVs. Intermediate results are streamed between stages using a producer–consumer mechanism, enabling on-the-fly consumption and improving cache locality. To maximize performance, ScalableHD integrates memory tiling and NUMA-aware worker-to-core binding. Further, it features two execution variants tailored for small and large batch sizes, each designed to exploit compute parallelism based on workload characteristics while mitigating the memory-bound compute pattern that limits HDC inference performance on modern multi-core CPUs. ScalableHD achieves up to $10\times$ speedup in throughput (samples per second) over state-of-the-art baselines such as TorchHD, across a diverse set of tasks ranging from human activity recognition to image classification, while preserving task accuracy. Furthermore, ScalableHD exhibits robust scalability: increasing the number of cores yields near-proportional throughput improvements. Together, these results establish ScalableHD as a general and practical solution for real-time, high-throughput deployment of HDC inference on multi-core CPU platforms.

% while retaining accuracy? (in the task part)

% with fixed encodings, a modular PyTorch-based framework, an end-to-end learned model

\end{abstract}


%%% Make sure you go through the bullets and refine them %%%

\section{Introduction}
\label{sec: intro}

Neural Network (NN) based Machine Learning (ML) and Deep Learning (DL) has revolutionized applications across several domains \cite{dla_survey}. Recently, transformer architecture \cite{vaswani2017attention} based Large Language Models (LLMs) have enabled unparalleled breakthroughs in diverse fields\textemdash ranging from Computer Vision (CV) and Natural Language Processing (NLP) to healthcare, biomedicine, banking and education \cite{llm_survey}. Despite the progress driven by NNs, NN models require significant computational resources, both for model training and inference \cite{trainableHD}. With the recent proliferation of available data and compute-intensive NN model deployment across applications, researchers have moved towards alternative paradigms of learning \cite{survey-hdc-stoch-framework}.

Hyperdimensional Computing (HDC) \cite{kanerva}, is a brain-inspired computing paradigm that has been gaining traction recently as such an alternative \cite{survey-hdc-stoch-framework}. HDC utilizes high-dimensional vectors, termed hypervectors (HVs), to enable highly parallel, noise-tolerant, and robust feature representation learning \cite{survey-two-parter-part-I}. In particular, HDC transforms raw input features into high-dimensional HVs, and performs model training on such encoded HVs to learn HV representations for each class of data in the training set \cite{kanerva}. Subsequently, during inference, a raw input feature vector is encoded to its representative HV, and assigned a class label based on its similarity (distance) in the hyperspace to the learned class HVs \cite{kanerva, gpu-imani-cascade-hd, rosing-vision-hd}. 

Despite HDCs benefits and success, traditional HDC frameworks perform poorly in terms of accuracy with respect to contemporary DL models \cite{trainableHD}. This poor accuracy performance can primarily be attributed to the single-pass, non-parametric learning methodology employed by traditional HDC \cite{trainableHD, trad-hdc-1, trad-hdc-2, trad-hdc-3-imani-voice-hd, trad-hdc-4-imani-rosing-hdna, trad-hdc-5-imani-rosing-hierarchical-hd, kanerva}. To address this issue, several works propose advanced encoding techniques to map raw input features to encoded HVs \cite{adv-encoding-1-hv-design-eff-hdc, adv-encoding-2-encoding-binarized-img, adv-encoding-3-hardware-aware-static-opt, adv-encoding-4-tiny-hd}. Complementary to encoding improvements, numerous works have explored adaptive learning of class HVs, drawing from DL techniques to bridge the accuracy gap in HDC frameworks \cite{static-encoding-1-programmable-hdc, static-encoding-2-seizure-det, static-encoding-3-colal-learning-secure-hdc, pim-dual-accl-cluster-pim, gpu-xcel-hd-efficient-gpu, static-encoding-6-genie-hd, static-encoding-7-efficient-human-activity-recognition-hdc-imani, static-encoding-8-distri-HD, static-encoding-9-hdc-framework-image-descriptors-cvpr, hetero-algorithm-hardware-co-design, static-encoding-11-imani-scalable-edge-hdc}. Notably, recent state-of-the-art (SOTA) frameworks have significantly improved HDCs accuracy performance by adaptively learning \emph{both} the HVs used to encode the raw input features, referred to as base HVs, as well as the class HVs \cite{trainableHD, gpu-xcel-hd-efficient-gpu, static-encoding-6-genie-hd}. The work proposed in \cite{trainableHD} further incorporates adaptive optimizer, dynamically adjusting the learning rate for updating base and class HVs, surpassing the accuracy performance of Deep NNs (DNNs) across several tasks. While HDC training, incorporating advanced learning mechanisms \cite{trainable-hd-old-version, trainableHD, multi-mani-hd-imani}, benefits from and often requires GPU acceleration due to the involved computational costs, performing inference with the trained base and class HVs retains the efficiency and parallelizability inherent to HDC.

Prior works \cite{survey-two-parter-part-II, survey-hdc-edge-intel-progress} in efficient HDC inference primarily focus on acceleration of HDC inference on platforms such as FPGAs \cite{fpga-e3-hdc, fpga-edge-ai-acc, fpga-general-hdc-edge, fpga-hdc-hardware-acc-cvpr, fpga-imani-revisiting-hdc-fpga, fpga-onlinehd, fpga-scalable-interpretable}, GPUs \cite{gpu-hdtorch, gpu-imani-cascade-hd, gpu-openhd, gpu-xcel-hd-efficient-gpu}, memory-centric architectures such as Processor-In-Memory (PIM) and Compute-In-Memory (CIM) \cite{pim-biohd-imani, pim-dual-accl-cluster-pim, pim-hdnn-rosing, pim-paap-hd} and heterogeneous architectures \cite{hetero-algorithm-hardware-co-design, hetero-hpvm-hdc, multi-comphd, multi-mani-hd-imani, multi-multiarch-hardware-acc, cpu-hdcc-compiler}. While works such as \cite{multi-comphd, multi-mani-hd-imani, multi-multiarch-hardware-acc, gpu-hdtorch} support deployment across multiple platforms, they are not optimized for multi-core CPUs. \cite{cpu-hdcc-compiler} introduces the HDCC compiler for portable C-based deployment of HDC models, but does not focus on high-throughput inference or scalable execution on modern multi-core CPU architectures.

Despite the ubiquity of multi-core CPUs in modern computing infrastructure, including high-performance workstations and multi-socket servers, prior work has not systematically addressed high-throughput and scalable HDC inference on such architectures. This represents a missed opportunity, as HDC inference relies on compute primitives such as bundling, binding, permutation, and similarity search that are inherently parallelizable across the HV dimensionality. However, recent advances in HDC frameworks rely on HVs with large dimensionality, to deliver high application accuracy \cite{survey-two-parter-part-I, survey-hdc-stoch-framework, survey-hdc-edge-intel-progress}, which significantly increases the data movement between memory and CPU cores, leading to memory-bound HDC inference performance. Naive parallelization on multi-core CPUs introduces additional bottlenecks associated with uncoordinated memory access patterns, poor cache reuse and imbalance in workload across cores, leading to poor inference performance and scalability \cite{advance-on-cpu-ref}.

Enabling high-throughput HDC inference is critical for real-time applications where large volumes of streaming data must be processed rapidly and consistently to support downstream decision-making \cite{survey-hdc-edge-intel-progress}. HDC has been successfully applied in continuous human activity monitoring, biosignal classification, and real-time emotion recognition \cite{static-encoding-7-efficient-human-activity-recognition-hdc-imani, pim-biohd-imani, static-encoding-2-seizure-det}, where inference must meet strict throughput demands imposed by incoming, real-time, sensor data streams. Achieving scalability is equally important: a scalable HDC inference algorithm ensures optimal utilization of available compute resources across diverse platforms and input sample sizes, enabling efficient inference under a wide range of conditions. Despite being widely available, accessible, and cost-effective, multi-core CPUs have received limited attention in prior work on HDC inference acceleration \cite{advance-on-cpu-ref, cpu-hdcc-compiler}, which has largely focused on specialized hardware such as FPGAs and GPUs \cite{survey-hdc-edge-intel-progress}. Realizing high-throughput and scalable HDC inference on general-purpose multi-core CPUs would enable HDC to serve as an efficient alternative to NN-based DL for real-time inference deployment\textemdash broadening its accessibility and applicability across a wide range of applications.

To address this gap, we propose ScalableHD\textemdash a novel two-stage pipelined execution model for high-throughput and scalable HDC inference on general-purpose multi-core CPUs. ScalableHD supports execution strategies for two distinct workload regimes: one optimized for efficient small-batch inference, and the other for high-throughput processing of large batches. To ensure scalability, ScalableHD incorporates memory tiling, NUMA-aware worker-to-core binding, and streaming of intermediate encoded HVs between pipeline stages\textemdash effectively mitigating the memory-bound performance bottlenecks of HDC inference. We summarize the contributions of our work here,

\begin{itemize}
    \item We propose ScalableHD\textemdash a novel method for high-throughput and scalable HDC inference on general-purpose multi-core CPU platforms.
    \item We propose two execution variants within ScalableHD: ScalableHD-S, optimized for inference on a small number of input samples, and ScalableHD-L, designed for high-throughput inference on large batch sizes.
    \item ScalableHD mitigates the memory-bound nature of HDC inference by implementing it as a two-stage pipeline, where intermediate high-dimensional HVs are streamed in block-partitioned tiles between stages.
    \item We further introduce optimizations, including memory tiling and NUMA-aware worker-to-core binding, to improve performance and enable scalable, high-throughput execution of ScalableHD on multi-core CPUs.
    \item We evaluate ScalableHD across diverse HDC tasks, showing up to $10\times$ higher inference throughput than CPU baselines like TorchHD, with no changes to the HDC algorithm or drop in accuracy.
    \item ScalableHD scales robustly on general-purpose multi-core CPUs, achieving near-linear throughput gains with increasing core counts. This demonstrates its generality and effectiveness.
\end{itemize} 

The remainder of this paper is organized as follows. Section~\ref{sec: prelim} introduces the necessary preliminaries for HDC, including basic HDC operations and the training setup. Section~\ref{sec: method} presents the proposed ScalableHD method, including its two-stage pipelined execution model, execution variants, and optimizations. Section~\ref{sec: exp} describes the experimental setup and evaluates ScalableHD across a range of datasets, CPU platforms and baselines. Section~\ref{sec: related} reviews related work on HDC inference acceleration. Finally, Section~\ref{sec: concl} concludes the paper and outlines directions for future work.

\section{Preliminaries}
\label{sec: prelim}
Hyperdimensional Computing (HDC), introduced in \cite{kanerva}, is a brain-inspired computing paradigm that encodes raw input features $\mathbf{x} \in \mathbb{R}^F$ into high-dimensional vectors $\mathbf{h} \in \mathbb{H}^D$ using a mapping $\mathcal{M}: \mathbb{R}^F \rightarrow \mathbb{H}^D$, where $\mathbb{H}^D$ denotes either the $D$-dimensional bipolar space $\{-1, +1\}^D$ or the binary space $\{0, 1\}^D$ \cite{kanerva, graph-hdgl}. Note that the typical value of $D$ is roughly $10^4$ for most HDC applications. Throughout this work, we assume bipolar encodings, with $\mathbb{H}^D = \{-1, +1\}^{D}$. 

An important characteristic of HDC HVs is the near-orthogonality of two randomly generated HVs \cite{kanerva, trainableHD}. Specifically, two randomly generated bipolar HVs, $\mathbf{h}_1, \mathbf{h}_2 \in \mathbb{H}^{D}$, will have an inner-product that is approximately zero, $\inner{\mathbf{h}_1}{\mathbf{h}_2} \approx 0$, for $D > 1000$ \cite{trainableHD, survey-two-parter-part-I}. Note, in our work, we will use inner-product as our similarity measure between two HVs. This near-orthogonality of randomly generated HVs enables the HDC learning paradigm to be noise-tolerant and robust to errors\cite{survey-hdc-stoch-framework}. While several encoding mechanisms have been proposed to define the mapping $\mathcal{M}$ \cite{trad-hdc-3-imani-voice-hd, static-encoding-3-colal-learning-secure-hdc, pim-dual-accl-cluster-pim, trainableHD}, this work adopts the advanced nonlinear encoding method introduced in \cite{pim-dual-accl-cluster-pim}. The nonlinear encoding defines the mapping $\mathcal{M}$ via a codebook of $F$ HVs, represented as a base HV matrix $\mathbf{B} \in \mathbb{R}^{F \times D}$, which transforms a raw input feature vector $\mathbf{x} \in \mathbb{R}^F$ into an encoded HV $\mathbf{h} \in \mathbb{H}^D$. After encoding, all computation in the HDC framework is performed over HVs in the high-dimensional hyperspace using a well-defined set of fundamental operations (compute primitives). We first describe these fundamental operations before detailing the nonlinear encoding process.

\subsection{HDC Operations}
\label{subsec: prelim_hdcops}

HDC relies on three fundamental operations: (i) bundling, (ii) binding, and (iii) permutation. Each operates element-wise over HVs and is inherently parallelizable. We describe each operation below.

\vspace{2pt}
\noindent \textbf{Bundling} We denote the bundling operation via $\bigoplus$. The bundling operation takes as input two HVs, $\mathbf{h}_1, \mathbf{h}_2 \in \mathbb{H}^{D}$, and performs element-wise addition along the HV dimensionality, $D$, aggregating the input HVs. The composite HV, $\mathbf{v} = \mathbf{h}_1 \bigoplus \mathbf{h}_2$, is a HV of size $D$, however $v \notin \mathbb{H}^D$. This is often referred to as unconstrained bundling \cite{survey-two-parter-part-I}, and needs to be normalized in order to return a HV in $\mathbb{H}^{D}$. For bipolar HVs, the normalization is often performed via a $\mathrm{HardSign()}$ function, defined below,

\begin{equation}
\label{eq: hardsign}
\mathrm{HardSign}(x) =
\begin{cases}
+1, & \text{if } x \geq 0 \\
-1, & \text{otherwise}
\end{cases}
\end{equation}

Note that the $\mathrm{HardSign()}$ function in eq. \ref{eq: hardsign}, is essentially a majority vote operation, with ties broken with $+1$. Applying this operation to the unnormalized output, $\mathbf{v}$, finally returns $\mathbf{h} = \mathrm{HardSign(\mathbf{v})} \in \mathbb{H}^{D}$. The bundling operation, like standard addition, is associative and commutative, and is utilized to aggregate information from the input HVs, yielding a resultant representative HV.   

\vspace{2pt}
\noindent \textbf{Binding} The binding operation is denoted via $\otimes$. Binding operation takes as input two HVs, $\mathbf{h}_1, \mathbf{h}_2 \in \mathbb{H}^{D}$, and performs element-wise multiplication along the HV dimensionality, \textit{binding} the input HVs. The resultant HV, $\mathbf{h} = \mathbf{h}_1 \otimes \mathbf{h}_2$ is a HV of dimension $D$, and $\mathbf{h} \in \mathbb{H}^{D}$. The binding operation is invertible, commutative, and associative. Note that $\mathbf{h} \otimes \mathbf{h}_1 = \mathbf{h}_2$ and $\mathbf{h} \otimes \mathbf{h}_2 = \mathbf{h}_1$, leading to invertibility. The binding operation is utilized to construct a HV that is near-orthogonal to both the input HVs, enabling the representation of a concept that is distinct from its constituents. We can also bind a real scalar, $c \in \mathbb{R}$, to a HV $\mathbf{h} \in \mathbb{H}^{D}$, by multiplying each element of $\mathbf{h}$ with the scalar $c$. The resultant HV, $\mathbf{v} = c \otimes \mathbf{h}$ is a HV of dimension $D$, $\mathbf{v} \notin \mathbb{H}^{D}$. Similar to bundling, $\mathbf{v}$ may be normalized for subsequent processing. 

\vspace{2pt}
\noindent \textbf{Permutation} The permutation operation, denoted by $\Pi^{(i)}$, takes as input a single HV $\mathbf{h} \in \mathbb{H}^D$ and cyclically rotates its elements by $i$ positions (left or right, used consistently throughout). The result, $\mathbf{h}' = \Pi^{(i)} \mathbf{h}$, is an HV in $\mathbb{H}^D$ that represents the rotated version of $\mathbf{h}$. The permutation (or rotation) operation is the HDC analogue of cyclic shifts in the binary domain and is primarily used to produce a HV that is dissimilar or near-orthogonal to the original.

We next describe the nonlinear encoding mechanism adopted in this work, followed by a discussion on HDC training and inference.

\subsection{Nonlinear Encoding}
\label{subsec: prelim_nonlinenc}
Nonlinear encoding \cite{pim-dual-accl-cluster-pim} is an advanced encoding mechanism that utilizes a codebook of base vectors, $\mathbf{B} \in \mathbb{R}^{F \times D}$, to encode the raw input feature vector, $\mathbf{x} \in \mathbb{R}^{F}$. The elements of the $F$ base vectors, $\mathbf{b}_1, \mathbf{b}_2, \ldots, \mathbf{b}_F$ (row vectors of matrix $\mathbf{B}$), are sampled from a Gaussian distribution. Considering the scalar elements of the raw input feature vector, $\mathbf{x}$, to be $x_1, x_2, \ldots, x_F$, the nonlinear encoding mechanism performs the below operation to encode $\mathbf{x}$ via the base matrix $\mathbf{B}$,

\begin{equation}
\label{eq: nonlinenc}
    \mathbf{v} = (x_1 \otimes \mathbf{b}_1) \oplus (x_2 \otimes \mathbf{b}_2) \oplus \ldots \oplus (x_F \otimes \mathbf{b}_F) 
\end{equation}

In eq. \ref{eq: nonlinenc}, the output HV $\mathbf{v}$ is computed by binding each input feature scalar to its corresponding base HV, followed by bundling the resulting bound vectors. The HV $\mathbf{v} \in \mathbb{R}^{D}$ is unnormalized. Nonlinear encoding typically applies a nonlinear activation to $\mathbf{v}$ for both normalization and increased representational expressivity. In our work, we use the $\mathrm{HardSign}()$ function described before, as the activation function. Thus,

\begin{equation}
\label{eq: nonlinenc_activ}
    \mathbf{h} = \mathrm{HardSign}(\mathbf{v})
\end{equation}

In eq. \ref{eq: nonlinenc_activ}, $\mathbf{h} \in \mathbb{H}^{D}$, is the final output of the nonlinear encoding, which is a bipolar HV.

\subsection{HDC Training}
\label{subsec: prelim_train}
Traditional HDC frameworks typically utilize a single-pass training algorithm. Given a dataset with $K$ classes, single-pass training constructs a class HV codebook $\mathbf{M} \in \mathbb{H}^{K\times D}$ by aggregating encoded HVs corresponding to each class. The aggregation is generally performed via normalized (constrained) bundling of the HVs belonging to the same class. Despite being computationally efficient, single-pass training leads to poor accuracy performance due to its inability to adapt the encoding to the data distribution and its non-parametric training \cite{trainableHD}.

Addressing these limitations, TrainableHD \cite{trainableHD} framework introduces a learnable approach to HDC training, inspired from DL, reaching state-of-the-art accuracy performance across several datasets. Specifically, TrainableHD jointly optimizes both: the base HV matrix and the class HV matrix, $\mathbf{B} \in \mathbb{R}^{F\times D}$ and $\mathbf{M} \in \mathbb{R}^{K\times D}$, respectively, using gradient-based optimization techniques. The training process utilizes an error vector, $\mathbf{e} \in \mathbb{R}^{K}$, that captures the discrepancy between the predicted and true class labels, to adaptively update the base and class HV matrix, $\mathbf{B}$ and $\mathbf{M}$, respectively. The adaptive updates for the learnable base and class HV matrix are performed via optimizers like Adam \cite{adam}, to optimize the learning rate ($\lambda$) during the end-to-end training process. 

For our work, we adopt the TrainableHD framework to learn the base and class HV matrices, $\mathbf{B}$ and $\mathbf{M}$, and use them directly in our two-stage ScalableHD inference pipeline to ensure accurate HDC inference. Both matrices are retained in 32-bit floating-point format. While TrainableHD supports quantization-aware training (QAT) to reduce model size and precision requirements \cite{qat-1, qat-2}, we avoid quantization to maintain inference accuracy and to leverage the high-throughput SIMD vectorization capabilities of modern CPUs, which are highly optimized for floating-point operations \cite{intel-simd}.

We next discuss HDC inference, using the (learned) base and class HV matrix.

\subsection{HDC Inference}
\label{subsec: prelim_inference}
Given a trained base HV matrix $\mathbf{B} \in \mathbb{R}^{F \times D}$ and class HV matrix $\mathbf{M} \in \mathbb{R}^{K \times D}$, HDC inference for a single input feature vector $\mathbf{x} \in \mathbb{R}^{F}$ proceeds as follows. First, nonlinear encoding is applied to compute the high-dimensional HV representation $\mathbf{h} \in \mathbb{H}^D$ of $\mathbf{x}$ using the method described in Section~\ref{subsec: prelim_nonlinenc}. Specifically, each scalar $x_i$ is bound to its corresponding base vector $\mathbf{b}_i$, and the results are bundled,

\begin{equation}
\label{eq: inf_single}
\mathbf{h} = \mathrm{HardSign} \left( \bigoplus_{i=1}^{F} (x_i \otimes \mathbf{b}_i) \right)
\end{equation}

The encoded HV $\mathbf{h}$ is then compared to each of the class HVs $\{\mathbf{m}_k\}_{k=1}^{K}$ (row vectors) in $\mathbf{M}$ via an inner product to compute similarity scores,

\begin{equation}
\label{eq: sim_single}
s_k = \inner{\mathbf{h}}{\mathbf{m}_k}, \quad \text{for } k = 1, \dots, K
\end{equation}

The final class label is determined by selecting the class with the highest similarity score,

\begin{equation}
\label{eq: pred_single}
y_{\text{pred}} = \arg\max_k \, s_k
\end{equation}

To efficiently handle a batch of $N$ input samples, we express HDC inference (described above) in matrix form. Let $\mathbf{X} \in \mathbb{R}^{N \times F}$ denote the input batch. The encoded HVs for the entire batch can then be computed as follows,

\begin{equation}
\label{eq: batch_encode}
\mathbf{H} = \mathrm{HardSign}(\mathbf{X} \mathbf{B}), \quad \mathbf{H} \in \mathbb{H}^{N \times D}
\end{equation}

Next, a similarity score matrix $\mathbf{S} \in \mathbb{R}^{N \times K}$ is computed by taking the inner product of each encoded HV with the $K$ class HVs, 

\begin{equation}
\label{eq: batch_sim}
\mathbf{S} = \mathbf{H} \mathbf{M}^\top
\end{equation}

Finally, the predicted class labels are obtained as,

\begin{equation}
\label{eq: pred_batch}
\mathbf{y}_{\text{pred}} = \arg\max (\mathbf{S}, \text{axis}=1)
\end{equation}

Note that the vector $\mathbf{y}_{\text{pred}}$ contains $N$ predictions for each sample. This end-to-end inference, naturally decomposes into two computational stages, described in Algorithm \ref{alg: hdc_inference}. Stage I computes the HV encoding of the input, whereas Stage II performs similarity search and classification, return the final predicted classes $\mathbf{y}_{\text{pred}}$. 

\begin{algorithm}
\caption{Two-Stage HDC Inference}
\label{alg: hdc_inference}
\begin{algorithmic}[1]
\Statex \textbf{Input:} $\mathbf{X} \in \mathbb{R}^{N \times F}$ (raw feature matrix), $\mathbf{B} \in \mathbb{R}^{F \times D}$ (base HV matrix), $\mathbf{M} \in \mathbb{R}^{K \times D}$ (class HV matrix)
\Statex \textbf{Output:} $\mathbf{y}_{\text{pred}} \in \{0, 1, \dots, K{-}1\}^N$

\State // Stage I: Feature Encoding
\State $\mathbf{H} \gets \mathrm{HardSign}(\mathbf{X}\mathbf{B})$ 

\State // Stage II: Similarity Search \& Classification
\State $\mathbf{S} \gets \mathbf{H}\mathbf{M}^\top$
\State $\mathbf{y}_{\text{pred}} \gets \arg\max(\mathbf{S}, \text{axis}=1)$
\State \Return $\mathbf{y}_{\text{pred}}$
\end{algorithmic}
\end{algorithm}

The two-stage inference process described above is applicable to all HDC methods that adopt the nonlinear encoding scheme \cite{pim-dual-accl-cluster-pim, trainableHD, static-encoding-7-efficient-human-activity-recognition-hdc-imani, gpu-xcel-hd-efficient-gpu}. Importantly, the intermediate matrix $\mathbf{H} \in \mathbb{H}^{N \times D}$, produced during Stage I, can become prohibitively large for high-dimensional settings and large input batches. This introduces significant memory bandwidth and data transfer bottlenecks, particularly when naively parallelized on general-purpose multi-core CPUs, severely limiting HDC inference throughput and scalability. We address these challenges through our proposed ScalableHD framework, introduced in the next section.

\section{Method}
\label{sec: method}

In this section, we present the ScalableHD method in detail. Section \ref{subsec: method_ovw} provides an overview of the two-stage pipelined design and related optimizations. Section \ref{subsec: method_sI} describes Stage I, which encodes raw input features into high-dimensional HVs. Section \ref{subsec: method_sII} covers Stage II, which performs similarity search and classification. Finally, Section \ref{subsec: method_opt} outlines system-level optimizations\textemdash including memory tiling and NUMA-aware worker-to-core binding\textemdash that improve ScalableHD's throughput and scalability.

\subsection{ScalableHD: Overview}
\label{subsec: method_ovw}
\begin{figure}
    \centering
    \includegraphics[width=0.75\linewidth]{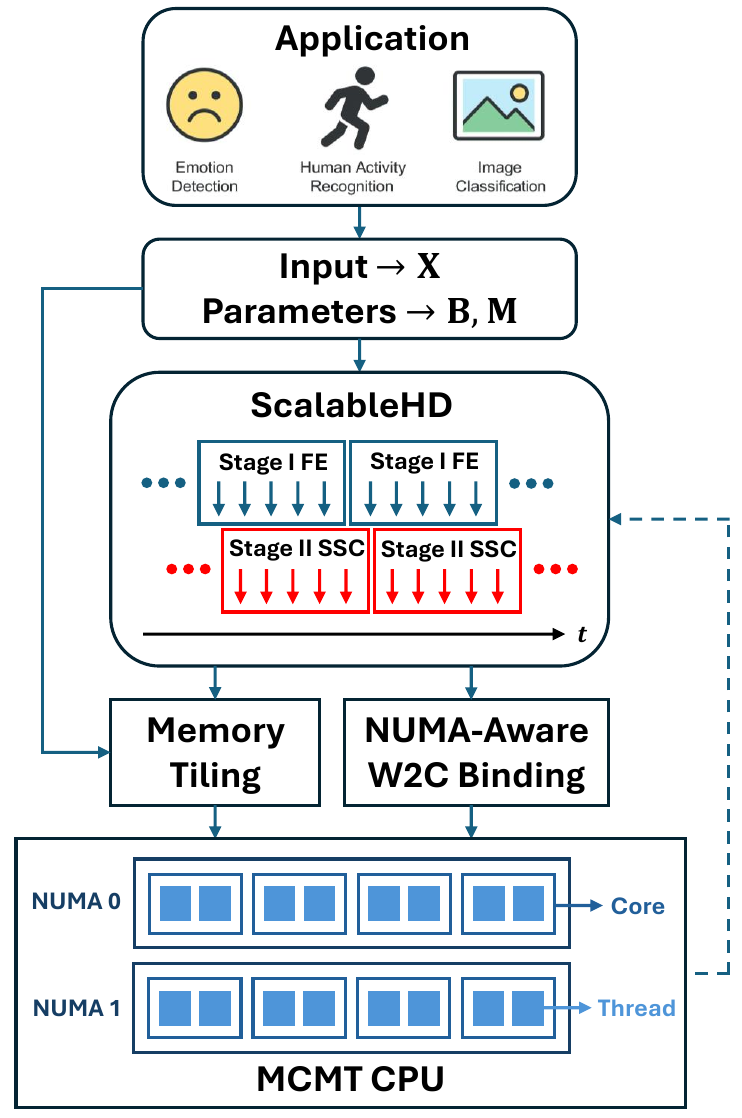}
    \caption{Overview of the proposed ScalableHD method. FE refers to Feature Encoding, SSC refers to Similarity Search \& Classification, and W2C refers to Worker-to-Core. MCMT refers to multi-core multi-threaded CPUs.}
    \label{fig: scalablehd_overview}
\end{figure}

Fig. \ref{fig: scalablehd_overview} presents the overview of ScalableHD and related optimizations. Given an application such as emotion detection \cite{emotion-det}, human activity recognition \cite{activity-rec-1, activity-rec-2}, and image classification \cite{class-3, class-4}, we utilize the TrainableHD \cite{trainableHD} framework for training, resulting in learned base and class matrices, $\mathbf{B} \in \mathbb{R}^{F\times D}$ and $\mathbf{M} \in \mathbb{R}^{K\times D}$. Note that ScalableHD is compatible with any HDC method that employs nonlinear encoding, including, but not limited to, \cite{trainableHD, pim-dual-accl-cluster-pim, static-encoding-7-efficient-human-activity-recognition-hdc-imani, gpu-xcel-hd-efficient-gpu}, making it broadly applicable across recent HDC frameworks.

Depending on the characteristics of the downstream application, the input batch $\mathbf{X} \in \mathbb{R}^{N\times F}$ may contain either a small or large number of samples, $N$, to be processed in parallel. ScalableHD includes two execution variants tailored for both settings: ScalableHD-S, optimized for small batches, and ScalableHD-L, designed for high-throughput processing of large batches. Both variants implement a two-stage pipelined algorithm. We refer to each parallel unit of execution in ScalableHD as a \emph{worker}.

\begin{itemize}
    \item Stage I: It performs the nonlinear encoding to transform the raw input feature matrix $\mathbf{X}$ into the encoded feature matrix $\mathbf{H}$, using the base matrix $\mathbf{B}$, via $\mathbf{H} = \mathrm{HardSign}(\mathbf{X}\mathbf{B})$. This stage is parallelized across $T$ workers, each responsible for computing a portion of $\mathbf{H}$. This stage is identical for both ScalableHD-S and ScalableHD-L; however, the two variants differ in how intermediate results are transferred to the workers of Stage II. Specifically, each variant adopts a distinct streaming and partitioning strategy for the portions of $\mathbf{H}$ computed by Stage I workers, optimized for either small or large batch processing.

    \item Stage II: This stage performs the similarity search and classification operation after feature encoding, by computing the matrix $\mathbf{S} = \mathbf{H}\mathbf{M}^\top$ and selecting the top scoring class for each input sample. Like Stage I, it is parallelized across $T$ workers; however, each variant employs a distinct execution model in Stage II. In ScalableHD-S, each worker computes a partial contribution to the full output matrix $\mathbf{S}$ by operating along the HV dimensionality ($D$). In contrast, ScalableHD-L assigns each worker a distinct subset of input samples ($N$) and computes the corresponding rows of $\mathbf{S}$. As encoded chunks of $\mathbf{H}$ are produced by Stage I, they are streamed to Stage II through lock-free queues, allowing workers in Stage II to begin computation as soon as data becomes available.
\end{itemize}

To further improve performance, ScalableHD incorporates two key system-level optimizations. First, workers in each stage are mapped to physical cores using NUMA-aware worker-to-core binding, which reduces non-local memory access and improves inter-stage data transfer latency. Second, the input matrix $\mathbf{X}$ and parameter matrices $\mathbf{B}$ and $\mathbf{M}$ are tiled in memory to enable contiguous access to matrix blocks during computation. This improves cache utilization, reducing memory access overhead for workers in each stage. Together, these optimizations enable ScalableHD to deliver high-throughput HDC inference across diverse applications and batch sizes, while scaling efficiently with increasing core counts.

\vspace{2pt}
\noindent \textbf{Notation} For any matrix $\mathbf{A} \in \mathbb{R}^{P \times Q}$, we define a block-wise view where $\mathbf{A}[i][j]$ denotes the $(i,j)$-th matrix block, with $0 \leq i < \lceil P/p \rceil$ and $0 \leq j < \lceil Q/q \rceil$, unless otherwise specified. Each block $\mathbf{A}[i][j]$ is a matrix of size $p \times q$, except for boundary blocks, which may be smaller. Specifically, $\mathbf{A}[i][j]$ refers to the block of matrix $\mathbf{A}$ containing the scalar elements $A_{\alpha, \beta}$, where $A_{\alpha, \beta}$ denotes the scalar value at row $\alpha$ and column $\beta$ of $\mathbf{A}$. The indices $(\alpha, \beta)$ for this block have the below range:
\[
ip \leq \alpha < \min((i + 1)p, P), \quad jq \leq \beta < \min((j+1)q, Q)
\]
We utilize this block-wise notation for the following matrices in the two-stage HDC inference pipeline (Algorithm \ref{alg: hdc_inference}): the input matrix $\mathbf{X} \in \mathbb{R}^{N \times F}$ is partitioned into blocks of size $n \times f$; the base matrix $\mathbf{B} \in \mathbb{R}^{F \times D}$ into blocks of size $f \times d$; the encoded matrix $\mathbf{H} \in \mathbb{R}^{N \times D}$ into blocks of size $n \times d$; the transposed class matrix $\mathbf{J} = \mathbf{M}^\top \in \mathbb{R}^{D \times K}$ into blocks of size $d \times k$; and the output similarity matrix $\mathbf{S} \in \mathbb{R}^{N \times K}$ into blocks of size $n \times k$. Note that we use $\mathbf{A}[i_1\!:\!i_2][j_1\!:\!j_2]$ to denote a block-sliced submatrix consisting of row blocks $i_1$ through $i_2-1$ and column blocks $j_1$ through $j_2-1$, and $\mathbf{A}(\alpha_1\!:\!\alpha_2,\ \beta_1\!:\!\beta_2)$ to denote an element-wise submatrix from rows $\alpha_1$ through $\alpha_2 - 1$ and columns $\beta_1$ through $\beta_2 - 1$. As shorthand, a colon symbol (:) denotes the full range in either dimension (e.g., $\mathbf{A}[i][:]$ or $\mathbf{A}(:,\beta)$). Additional memory-tiling and data layout optimizations are described in Section~\ref{subsec: method_opt}.

\subsection{Stage I: Feature Encoding}
\label{subsec: method_sI}

\begin{figure}
    \centering
    \includegraphics[width=\linewidth]{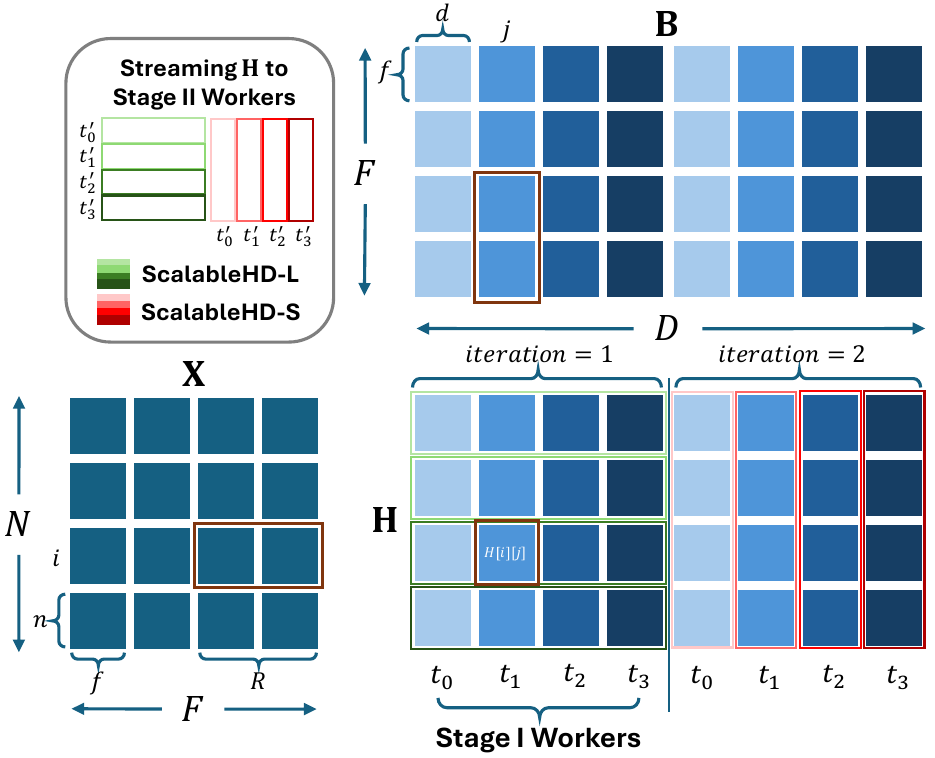}
    \caption{Stage I. Each worker computes a subset of output column blocks of $\mathbf{H}$ (e.g., worker $t_2$ computes column blocks 2 and 6). In ScalableHD-S (red), each worker streams its computed column blocks to the corresponding Stage II worker (e.g., $t_2 \rightarrow t_2'$). In ScalableHD-L (green), each worker streams $N/T$ disjoint rows of each column block to all Stage II workers (e.g., the first $N/T$ rows to $t_0'$).}
    \label{fig: stage_1}
\end{figure}

Fig.~\ref{fig: stage_1} shows Stage~I (Feature Encoding) of ScalableHD with the corresponding algorithm described in Algorithm~\ref{alg:stage1}. Each of the $T$ workers in Stage~I is responsible for computing a subset of output column blocks of the encoded matrix $\mathbf{H}$. Specifically, worker $t$ computes column blocks indexed by $t, t + T, t + 2T, \ldots$, of matrix $\mathbf{H}$. For each assigned column block $j$, the worker initializes an accumulation buffer $\mathbf{H}[:][j] \in \mathbb{R}^{N \times d}$ to store intermediate results. The final value at position $(i, j)$ of $\mathbf{H}$ is computed as a sum of partials, accumulated into $\mathbf{H}[i][j]$ in the buffer (line~\ref{algline: acc} of Algorithm~\ref{alg:stage1}). Each such partial is obtained by multiplying $R$ blocks in row block $i$ of matrix $\mathbf{X}$ with the corresponding $R$ blocks in column block $j$ of matrix $\mathbf{B}$ (see Fig.~\ref{fig: stage_1}). The algorithm iterates over row blocks of $\mathbf{X}$, reusing the same $R$ blocks of $\mathbf{B}$ to accumulate partials into each corresponding row block of the buffer $\mathbf{H}[:][j]$.

\begin{algorithm}
\caption{Stage I: Feature Encoding}
\label{alg:stage1}
\begin{algorithmic}[1]
\Statex \textbf{Input:} $\mathbf{X} \in \mathbb{R}^{N \times F}$, $\mathbf{B} \in \mathbb{R}^{F \times D}$
\Statex \textbf{Output:} $\mathbf{H} \in \mathbb{R}^{N \times D}$ (streamed to stage II workers)

%%%%%%%%%%%% specify tiling in the content %%%%%%%%%%%%
%\State Assume all matrices are partitioned into uniform tiles of size $n \times f$, $f \times d$, $n \times d$, respectively.

\State $T \gets$ the number of workers in stage I

\For{each worker $t = 0$ to $T - 1$ \textbf{in parallel}}
    \Statex // worker $t$ processes column blocks $j$ s.t. $j \bmod T = t$
    \For{each assigned column block $j$ of $\mathbf{H}$}
        \State initialize accumulation buffer $\mathbf{H}[:][j] \gets \mathbf{0}$
        \Statex // work on $R$ blocks in a row of $\mathbf{X}$ at a time
        \For{$r = 0, R, 2R, \ldots, \lceil F/f \rceil - 1$}
            \State $r'\gets \min(r + R - 1,\ \lceil F/f \rceil - 1)$
            \For{each row block $i$ of $\mathbf{X}$}
                \State $\mathbf{h}_{\text{local}} \gets \mathbf{0}$
                \For{column block $k$ in $r$ to $r'$}
                    \State $\mathbf{h}_{\text{local}} \mathrel{+}= \mathbf{X}[i][k] \mathbf{B}[k][j]$
                \EndFor
                \State $\mathbf{H}[i][j] \mathrel{+}= \mathbf{h}_{\text{local}}$ \label{algline: acc}
            \EndFor
        \EndFor
        \State $\mathbf{H}[:][j] \gets \mathrm{HardSign}(\mathbf{H}[:][j])$
        \Statex // streaming to stage II workers
        \If{ScalableHD-L}
            \For{each stage II worker $t'$}
                \State send $\mathbf{H}(t'\frac{N}{T}\!:\!(t'+1)\frac{N}{T},jd\!:\!(j+1)d)$ to queue $Q_{t'}$
            \EndFor
        \Else \Comment{ScalableHD-S}
            \State send $\mathbf{H}[:][j]$ to queue $Q_t$ of stage II worker $t$
        \EndIf
    \EndFor
    \State mark stage I worker $t$ as done
\EndFor
\end{algorithmic}
\end{algorithm}

This block-wise computation improves cache utilization and enables data reuse. Specifically, reusing the same column blocks of $\mathbf{B}$ across multiple row blocks of $\mathbf{X}$ within a compute round enhances data locality and reduces memory access overhead. Finally, the $\mathrm{HardSign()}$ operation is applied to the accumulated column block, which is then streamed to Stage II workers via lock-free queues. This Stage I procedure is identical for both variants of ScalableHD, with the only difference being in how the computed output column block $\mathbf{H}[:][j]$ is streamed to Stage II workers. In ScalableHD-S, the entire column block computed by Stage I worker $t$ is sent to corresponding Stage II worker $t$. In contrast, ScalableHD-L partitions the column block row-wise, and worker $t$ sends $N/T$ disjoint rows to each Stage II worker. Thus, ScalableHD-S uses one-to-one communication between Stage I and Stage II workers, while ScalableHD-L adopts an all-to-all communication pattern.

\subsection{Stage II: Similarity Search \& Classification}
\label{subsec: method_sII}

Stage II differs between the two variants of ScalableHD, each optimized for different input batch sizes. ScalableHD-S targets scenarios with a small number of input samples $N$, while ScalableHD-L is designed for high-throughput processing of large batches. Consequently, the two variants adopt distinct parallelization strategies. ScalableHD-S parallelizes computation along the HV dimensionality $D$ of the encoded matrix $\mathbf{H}$, whereas ScalableHD-L parallelizes across input samples $N$. We now describe the execution strategy for Stage II under both variants.

\begin{algorithm}
\caption{Stage II: Similarity Search (ScalableHD-S)}
\label{alg:stage2_s}
\begin{algorithmic}[1]
\Statex \textbf{Input:} Column blocks $\mathbf{H}[:][j]$ from stage I, $\mathbf{J} \in \mathbb{R}^{D \times K}$
\Statex \textbf{Output:} $\mathbf{S} \in \mathbb{R}^{N \times K}$, $\mathbf{y}_{\text{pred}} \in \{0, 1, \ldots, K-1\}^{N}$

\For{each worker $t' = 0$ to $T - 1$ \textbf{in parallel}}
    \State initialize local output buffer $\mathbf{S}_{\text{local}} \gets \mathbf{0}$
    \While{$Q_{t'}$ not empty \textbf{or} stage I worker $t'$ not done}
        \State try to dequeue column block $\mathbf{H}[:][j]$ from $Q_{t'}$
        \If{dequeue fails}
            \State \textbf{continue}
        \EndIf
        \For{each row block $i$ of $\mathbf{H}[:][j]$}
            \For{each column block $k$ of $\mathbf{J}[j][:]$}
                \State $\mathbf{S}_{\text{local}}[i][k] \mathrel{+}= \mathbf{H}[i][j] \mathbf{J}[j][k]$
            \EndFor
        \EndFor
    \EndWhile
    \Statex // accumulate local buffer into global matrix
    \State $\mathbf{S} \mathrel{+}= \mathbf{S}_{\text{local}}$ \Comment{critical section}
\EndFor

\Statex // final label prediction
\For{each sample $i = 0$ to $N - 1$ \textbf{in parallel}} 
    \State $\mathbf{y}_{\text{pred}}(i) \gets \arg\max_k \mathbf{S}(i,k)$
\EndFor
\end{algorithmic}
\end{algorithm}

\begin{figure}
    \centering
    \includegraphics[width=0.8\linewidth]{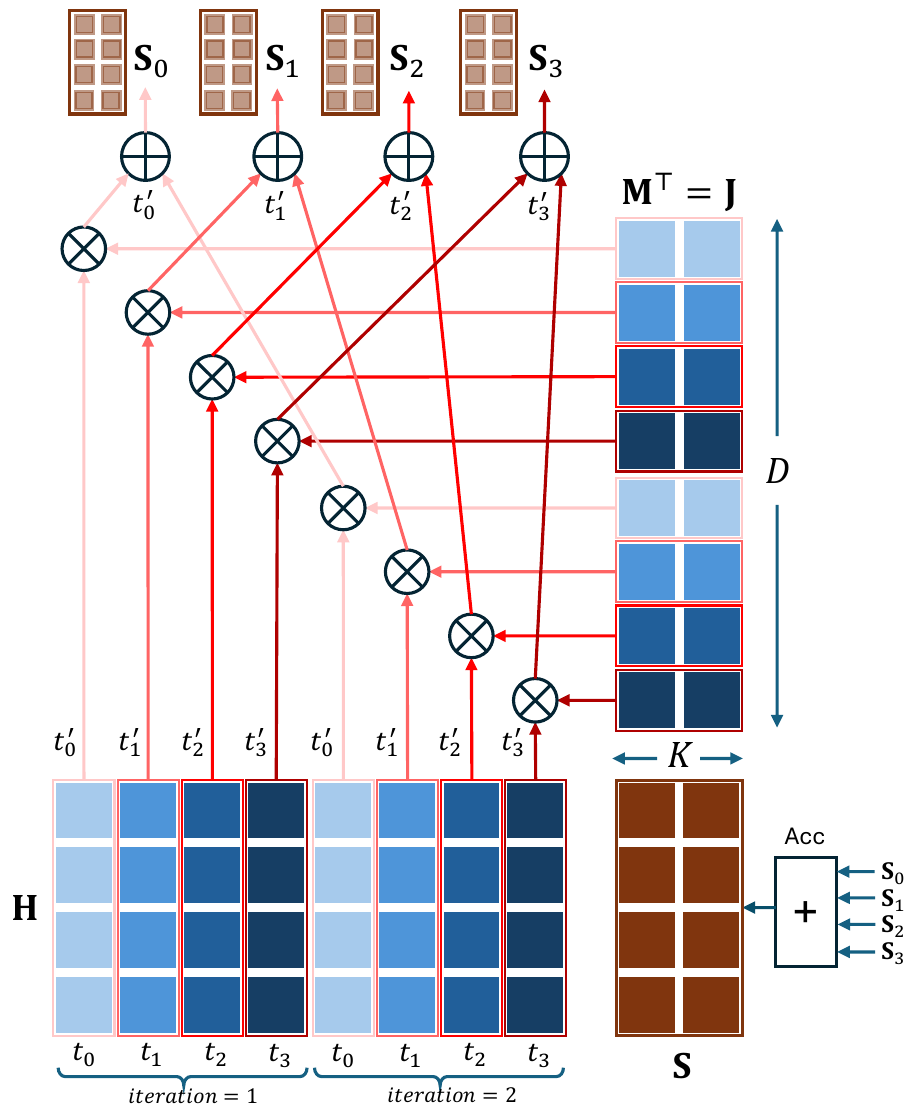}
    \caption{Stage~II of ScalableHD-S. Worker $t_0$ in Stage~I computes column blocks $0$ and $4$ of $\mathbf{H}$ and streams them to its sibling worker $t_0'$ in Stage~II. Worker $t_0'$ computes a partial of the full output matrix $\mathbf{S}$ (shown as $\mathbf{S}_0$) by multiplying each column block with the corresponding row block of $\mathbf{J}$.}
    \label{fig: stage2_s}
\end{figure}

\vspace{2pt}
\noindent \textbf{ScalableHD-S} In ScalableHD-S (Algorithm \ref{alg:stage2_s}, Fig. \ref{fig: stage2_s}), each Stage~II worker $t'$ checks its lock-free task queue $Q_{t'}$ for encoded column blocks $\mathbf{H}[:][j]$ produced by its sibling Stage~I worker $t'$. The worker continues dequeuing until the queue is empty \emph{and} its corresponding Stage~I worker has completed streaming all assigned column blocks of $\mathbf{H}$. Each dequeued block $\mathbf{H}[:][j]$, where $j \bmod T = t'$, is multiplied with the corresponding row block $\mathbf{J}[j][:]$ of the transposed class matrix, producing partial results that are accumulated into a local buffer $\mathbf{S}_{\text{local}} \in \mathbb{R}^{N \times K}$. This parallelization strategy distributes computation along the HV dimensionality $D$, allowing each worker to compute a partial result of the final output matrix $\mathbf{S}$. While ScalableHD-S requires $T$ local buffers with a total memory footprint of $NKT$, the small values of $N$ (in ScalableHD-S) and $K$ (typical for HDC) make this overhead manageable. By parallelizing along the HV dimensionality, ScalableHD-S enables fine-grained parallelism and improves per-worker compute utilization. The streamed $\mathbf{H}$ blocks of size $N\times d$ enhance cache locality and reduce memory access latency to the high-dimensional $\mathbf{H}$ matrix, further aided by system-level optimizations (see Section~\ref{subsec: method_opt}). Finally, each worker adds its local buffer to the global matrix $\mathbf{S}$, followed by a parallel $\arg\max$ operation across rows to produce the final predictions $\mathbf{y}_{\text{pred}}$.

\begin{figure}
    \centering
    \includegraphics[width=0.8\linewidth]{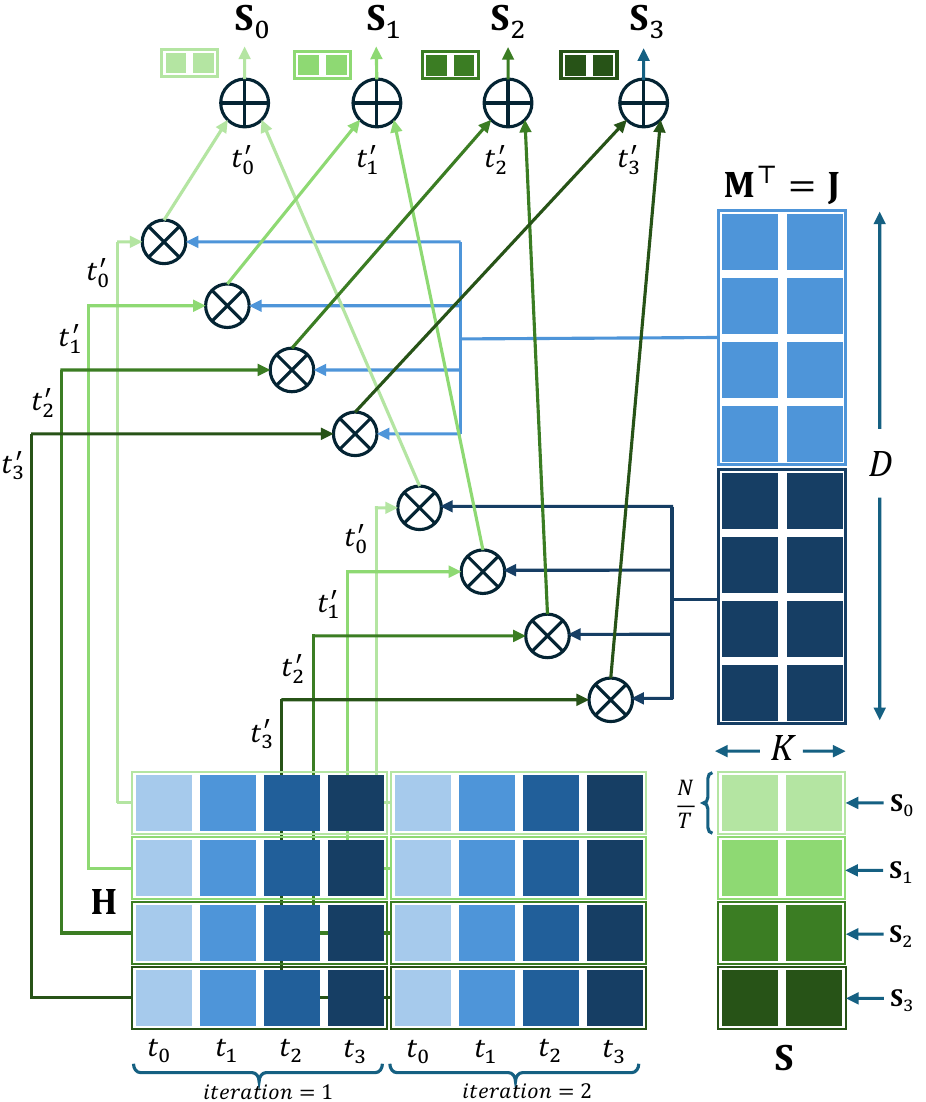}
    \caption{Stage II of ScalableHD-L. All Stage I workers ($T=4; t_0, \ldots, t_3$) stream the first $N/T$ rows to Stage II worker $t_0'$, which computes the first $N/T$ rows of the output matrix $\mathbf{S}$ (shown as $\mathbf{S}_0$).}
    \label{fig: stage2_l}
\end{figure}

\begin{algorithm}
\caption{Stage II: Similarity Search (ScalableHD-L)}
\label{alg:stage2_l}
\begin{algorithmic}[1]
\Statex \textbf{Input:}  Chunks of column block $\mathbf{H}(t'\frac{N}{T}\!:\!(t'+1)\frac{N}{T},jd\!:\!(j+1)d)$ from stage I, $\mathbf{J} \in \mathbb{R}^{D \times K}$
\Statex \textbf{Output:} $\mathbf{S}^{(t')} \in \mathbb{R}^{\frac{N}{T} \times K}$, $\mathbf{y}_{\text{pred}}^{(t')} \in \{0,1,\ldots,K{-}1\}^{\frac{N}{T}}$

\Statex // $\mathbf{H}^{(t', j)}$ refers to $\mathbf{H}(t'\frac{N}{T}\!:\!(t'+1)\frac{N}{T},jd\!:\!(j+1)d)$

\For{each stage II worker $t' = 0$ to $T - 1$ \textbf{in parallel}}
    \State initialize local output buffer $\mathbf{S}^{(t')} \gets \mathbf{0}$
    \While{$Q_{t'}$ not empty \textbf{or} any stage I worker not done}
        \State try to dequeue $\mathbf{H}^{(t', j)}$ from $Q_{t'}$
        \If{dequeue fails}
            \State \textbf{continue}
        \EndIf
        \For{each row block $i$ of $\mathbf{H}^{(t', j)}$}
            \For{each column block $k$ of $\mathbf{J}[j][:]$}
                \State $\mathbf{S}^{(t')}[i][k] \mathrel{+}= \mathbf{H}^{(t', j)}[i][j] \mathbf{J}[j][k]$
            \EndFor
        \EndFor
    \EndWhile
    \For{each row $i$ in $\mathbf{S}^{(t')}$}
        \State $\mathbf{y}_{\text{pred}}^{(t')}(i) \gets \arg\max_k \mathbf{S}^{(t')}(i,k)$
    \EndFor
    \State write $\mathbf{y}_{\text{pred}}^{(t')}$ to global $\mathbf{y}_{\text{pred}}$
\EndFor
\end{algorithmic}
\end{algorithm}

\vspace{2pt}
\noindent \textbf{ScalableHD-L} In ScalableHD-L (Algorithm \ref{alg:stage2_l}, Fig. \ref{fig: stage2_l}), to support large input batch sizes $N$, each Stage~I worker streams $N/T$ disjoint rows of its computed column block $j$ to every Stage~II worker $t'$. As a result, each Stage~II worker $t'$ receives a full chunk of $N/T$ rows of the encoded matrix $\mathbf{H}$, composed of contributions from all column blocks $j$ (see Fig. \ref{fig: stage2_l}). Thus, each Stage~II worker $t'$ is responsible for computing a disjoint set of $N/T$ rows of the output matrix $\mathbf{S}$, stored in its local buffer $\mathbf{S}^{(t')} \in \mathbb{R}^{\frac{N}{T} \times K}$. For each received column block chunk $\mathbf{H}^{(t', j)}$, the worker multiplies it with the corresponding row block $\mathbf{J}[j][:]$ of the transposed class matrix to compute partial results for $\mathbf{S}^{(t')}$. These partials are accumulated across all column blocks to complete the computation of the $N/T$ rows of $\mathbf{S}$ assigned to worker $t'$. After its task queue $Q_{t'}$ is empty \emph{and} all Stage~I workers are marked done, the worker performs a row-wise $\arg\max$ over $\mathbf{S}^{(t')}$ to produce a portion of the prediction vector $\mathbf{y}_{\text{pred}}^{(t')}$, which is written out to the global prediction vector. This variant parallelizes computation along the sample dimension $N$, enabling fine-grained parallelism across workers for large input batches. Unlike ScalableHD-S, the memory footprint remains fixed at $NK$, independent of $T$. However, ScalableHD-L introduces an all-to-all communication pattern between Stage~I and Stage~II workers: each of the $T$ Stage~I workers streams $\frac{N}{T} \times d$-sized tiles of $\mathbf{H}$ to every Stage~II worker. Thus, while both variants communicate a total of $NdT$ elements of $\mathbf{H}$ per round, ScalableHD-S performs one-to-one communication between sibling workers, whereas ScalableHD-L performs all-to-all communication. The block-partitioned streams improve cache locality and promote data reuse; however, inter-core bandwidth constraints and contention may limit overall performance, which is mitigated by our system-level optimizations (see Section~\ref{subsec: method_opt}).

\subsection{Optimizations}
\label{subsec: method_opt}

In this section, we present two key system-level optimizations—memory tiling and NUMA-aware worker-to-core binding—that enable the scalable and high-throughput inference performance of ScalableHD.

\begin{figure}
    \centering
    \includegraphics[width=\linewidth]{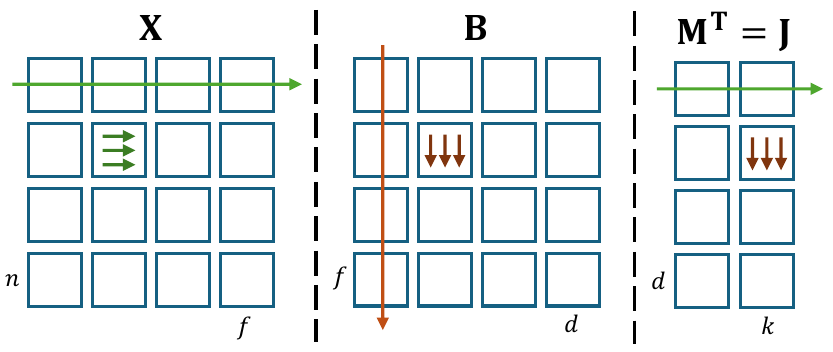}
    \caption{Memory tiling of $\mathbf{X}$, $\mathbf{B}$, and $\mathbf{J}$. Green (red) arrows indicate row-major (column-major) layout at both the inter-block and intra-block (elements within a block) level.}
    \label{fig: memory_tiling}
\end{figure}

\vspace{2pt}
\noindent \textbf{Memory Tiling} We tile the input matrix $\mathbf{X}$ and parameter matrices $\mathbf{B}$ and $\mathbf{J} = \mathbf{M}^\top$ to align with the block-wise access patterns in Stage~I and Stage~II computations (see Fig.~\ref{fig: memory_tiling}). During Stage~I, each worker computes a column block of $\mathbf{H}$ by multiplying $R$ blocks in a row block of $\mathbf{X}$ with the corresponding $R$ blocks in a column block of $\mathbf{B}$, during a given compute round. To enable contiguous memory access during Stage~I, $\mathbf{X}$ is tiled such that blocks are laid out in row-major order, with the elements within each block also stored in row-major order. In contrast, $\mathbf{B}$ is tiled with blocks arranged in column-major order, with the elements within each block stored in column-major order. These layouts align with the access patterns during block-wise matrix multiplication of $\mathbf{X}$ (left matrix) and $\mathbf{B}$ (right matrix), enabling efficient streaming memory access and improved cache utilization.

Each computed column block of $\mathbf{H}$ in Stage~I is stored in row-major format, with all its blocks naturally contiguous in memory. This layout enables efficient streaming access to blocks within a column block of $\mathbf{H}$ during Stage~II computation. In Stage~II, each column block of $\mathbf{H}$ (or its chunk) is multiplied with the corresponding row block of $\mathbf{J} = \mathbf{M}^\top$. To support this access pattern, the blocks of $\mathbf{J}$ are arranged in row-major order, and the elements within each block are stored in column-major order, consistent with its use as the right matrix in block-wise matrix multiplication. The local buffers for $\mathbf{S}$, computed by each Stage~II worker, are stored in standard row-major layout to enable efficient row-wise $\arg\max$ operations for computing the final predictions $\mathbf{y}_{\text{pred}}$.

\begin{figure}
    \centering
    \includegraphics[width=\linewidth]{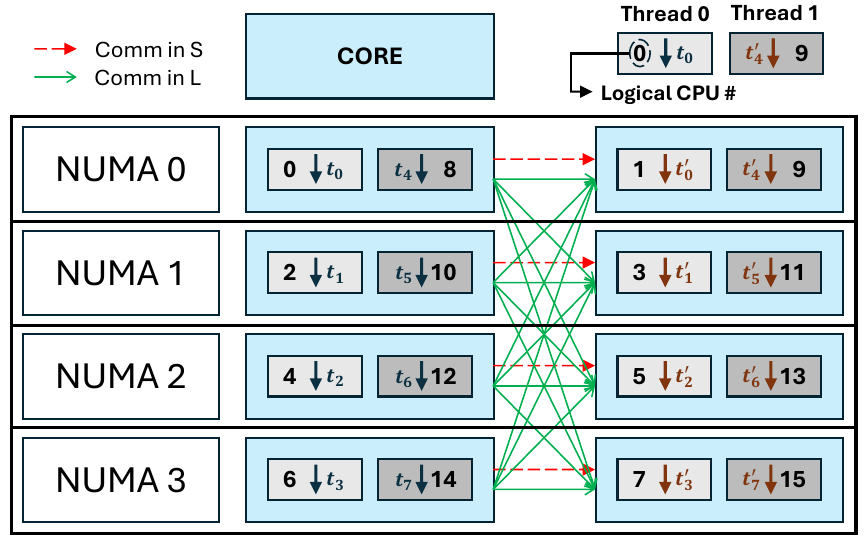}
    \caption{NUMA-aware worker-to-core binding. Example shown for $\gamma {=} 8$ cores and $\eta {=} 4$ NUMA nodes with 2 threads per core. Logical CPUs 0-7 correspond to thread~0; 8-15 to thread~1. For $T{=}4$, all workers are assigned to thread~0; for $T{=}8$, thread~1 is also used. Red (ScalableHD-S) and green (ScalableHD-L) lines indicate one-to-one and all-to-all communication patterns, respectively.}
    \label{fig: numa_bind}
\end{figure}

\vspace{2pt}
\noindent \textbf{NUMA-Aware Worker-to-Core Binding} Modern multi-core CPUs generally follow a Non-Uniform Memory Access (NUMA) architecture, where cores within a NUMA node can communicate via high-speed interconnects, leading to low-latency access to their shared memory pool. To mitigate the memory access overheads during inference, ScalableHD adopts a NUMA-aware worker-to-core binding strategy that co-locates communicating workers of the two-stage pipeline on the same NUMA node whenever possible. We consider a system with $\gamma$ physical cores and $\eta$ NUMA nodes, such that each node contains $\frac{\gamma}{\eta}$ cores. We assume Simultaneous Multi-Threading (SMT) with two hardware threads per core, a typical setup, giving us a total of $2\gamma$ logical CPUs. For NUMA node $\ell$ $(0 \leq \ell < \eta)$, thread 0 of all the cores on this node have logical CPU indices ranging from $\ell \left(\frac{\gamma}{\eta}\right)$ to $\left(\ell + 1\right)\left(\frac{\gamma}{\eta}\right) - 1$. Thread 1 of the cores, on the other hand, have indices ranging from $\gamma + \ell \left(\frac{\gamma}{\eta}\right)$ to $\gamma + \left(\ell + 1\right)\left(\frac{\gamma}{\eta}\right) - 1$. In Fig. \ref{fig: numa_bind}, $\gamma = 8$ and $\eta = 4$. We assign the total $2T$ workers ($T$ each for Stage~I and Stage~II) to these logical CPUs under the constraint $2T \leq 2\gamma$. The mapping first assigns workers to thread~0 of each core, exploiting core-level parallelism and maximizing resource utilization. At NUMA node $\ell$, the thread~0 logical CPUs span the range $\ell \left(\frac{\gamma}{\eta}\right)$ to $\left(\ell + 1\right)\left(\frac{\gamma}{\eta}\right) - 1$. Within this range, logical CPUs with even indices are assigned to Stage~I workers, and those with odd indices are assigned to the corresponding Stage~II workers. This assignment proceeds along the NUMA nodes in a similar manner until all thread~0 cores are utilized. If additional workers remain, a second pass begins, assigning them to thread~1 of each core following the same mapping strategy. In ScalableHD-S, this strategy guarantees that each Stage~I and Stage~II worker pair that communicates resides on the same NUMA node, minimizing communication latency. For ScalableHD-L, where all Stage~I workers interact with all Stage~II workers, the mapping maximizes the number of interacting pairs on the same NUMA node---improving cache locality and reducing memory access latency and contention.

\section{Experiments}
\label{sec: exp}
We conduct experiments with ScalableHD across diverse datasets and two state-of-the-art multi-core CPU platforms to validate its throughput and scalability performance. This section details the datasets used, hardware platforms, experimental settings, throughput and scalability analysis, and ablation results for ScalableHD.

\subsection{Datasets}
\label{subsec: dat}

\begin{table}
    \centering
    \caption{Datasets used for evaluation.}
    \label{tab: datasets}
    \renewcommand{\arraystretch}{1.1}
    \begin{adjustbox}{max width=0.5\textwidth}
    \begin{tabular}{llccccc}
        \toprule
        \textbf{Dataset} & \textbf{Task} & $F$ & $K$  & \# \textbf{Train} & \# \textbf{Test} & \textbf{Accuracy (\%)} \\
        \midrule
        MNIST\cite{class-3}  & Image Classification & 784   & 10  & 60,000 & 10,000 & 97.5 \\
        TEX\cite{class-4}   & Plant Classification  & 64   & 100       & 1,439  & 160    & 80.6 \\
        
        PAMAP2\cite{activity-rec-1}  & Activity Monitoring  & 27    & 5      & 16,384 & 16,384 & 92.4 \\
        HACT\cite{hact}     & Human Activity Recognition & 1152  & 6   & 7,352 & 2,947  & 85.8  \\
        SA12\cite{activity-rec-2}   & Smartphone Activity Recognition   & 561   & 12    & 6,213  & 1,554  & 97.8 \\
         ISOLET\cite{hact}  & Voice Recognition & 617   & 26         & 6,238  & 1,559  &  95.8 \\
        EMOTION\cite{emotion-det}  & Emotion Recognition & 1500  & 3        & 1,705  & 427    &  91.1 \\
        HEART\cite{ecg-heart}   & Arrhythmia Classification & 187   & 5    & 119,560 & 4,000 & 92.8 \\
        \bottomrule
    \end{tabular}
    \end{adjustbox}
\end{table}

For our experiments, we utilize a total of eight datasets spanning tasks such as image classification \cite{class-3, class-4}, human activity recognition \cite{activity-rec-1, activity-rec-2, hact}, and classification on bio-signals (such as ECG) \cite{emotion-det, ecg-heart}. Table \ref{tab: datasets} summarizes key statistics for each dataset. All datasets were trained using the TrainableHD framework \cite{trainable-hd-old-version, trainableHD}, with $D=10000$ (see Section \ref{subsec: setting} for details on training).

\subsection{Platforms}
\label{subsec: plat}

We evaluate ScalableHD on two state-of-the-art multi-core CPU platforms: a dual-socket AMD EPYC 7313 16-core processor (\ding{72}) and a dual-socket Intel Xeon Gold 6530 32-core processor (\ding{171}), with a total of 32 and 64 physical cores, respectively. Table \ref{tab: platform-specs} summarizes the hardware specifications, including core count, total NUMA nodes, cache size, peak memory bandwidth, and operating frequency.

\begin{table}[t]
    \centering
    \caption{CPU platform specifications used for evaluation.}
    \label{tab: platform-specs}
    \begin{adjustbox}{max width=0.4\textwidth}
    \renewcommand{\arraystretch}{1.2}
    \begin{tabular}{lcc}
        \toprule
         & \textbf{AMD EPYC 7313 (\ding{72})} & \textbf{Intel Xeon Gold 6530 (\ding{171})} \\
        \midrule
        Architecture & x86\_64 & x86\_64 \\
        Sockets & 2 & 2 \\
        Cores Per Socket & 16 & 32 \\
        Threads Per Core & 2 & 2 \\
        Total Cores & 32 & 64 \\
        %Total Threads & 64 & 128 \\
        NUMA Nodes & 2 & 4 \\
        Base Frequency & 3 GHz & 2.1 GHz \\
        Max Frequency & 3.7 GHz & 4 GHz \\
        L3 Cache (Total) & 256 MB & 320 MB \\
        %Peak Performance & & \\
        Peak Memory Bandwidth & 410 GB/s & 615 GB/s \\
        
        %Virtualization & AMD-V & VT-x \\
        \bottomrule
    \end{tabular}
    \end{adjustbox}
\end{table}

\subsection{Experimental Setting}
\label{subsec: setting}

\vspace{2pt}
\noindent \textbf{Training} We use the TrainableHD framework \cite{trainableHD} to train an HDC model for each dataset listed in Table~\ref{tab: datasets}. All models use a fixed HV dimensionality of $D=10000$ and are trained for $50$ epochs. Training is conducted on a single NVIDIA RTX 6000 Ada GPU using a batch size of $32$. We employ the Adam optimizer \cite{adam}, following the settings in \cite{trainableHD}, with an initial learning rate of $10^{-4}$. We do not use QAT, and the learned parameter matrices, $\mathbf{B}$ and $\mathbf{M}^\top = \mathbf{J}$, are retained in 32-bit floating-point format. This design choice enables efficient SIMD vectorization on multi-core CPUs, which are optimized for floating-point arithmetic, while also preserving accuracy by avoiding quantization. The trained $\mathbf{B}$ and $\mathbf{J}$ matrices are then used for inference on their respective datasets. We report the test accuracies in Table~\ref{tab: datasets}.

\vspace{2pt}
\noindent \textbf{Inference} We implement ScalableHD inference in C++ using POSIX threads (pthreads) \cite{pthreads} and OpenMP \cite{openmp} to exploit multi-core parallelism, \texttt{moodycamel::ConcurrentQueue} \cite{concurrentqueue} for lock-free task queues enabling efficient inter-stage communication, and Eigen \cite{eigen} for high-performance SIMD-accelerated matrix operations. We heuristically select the hyperparameters for ScalableHD algorithm. Specifically, we use standard tile sizes, $n = f = d = k \in \{16, 32\}$, to promote efficient use of first-level cache. The number of tiles processed per round in Stage I, $R \in \{8, 16\}$, is selected to enable data reuse from last-level cache. 

\begin{figure*}
    \centering
    % ========== AMD Plots ==========
    \begin{subfigure}[t]{0.48\textwidth}
        \centering
        \includegraphics[width=\linewidth]{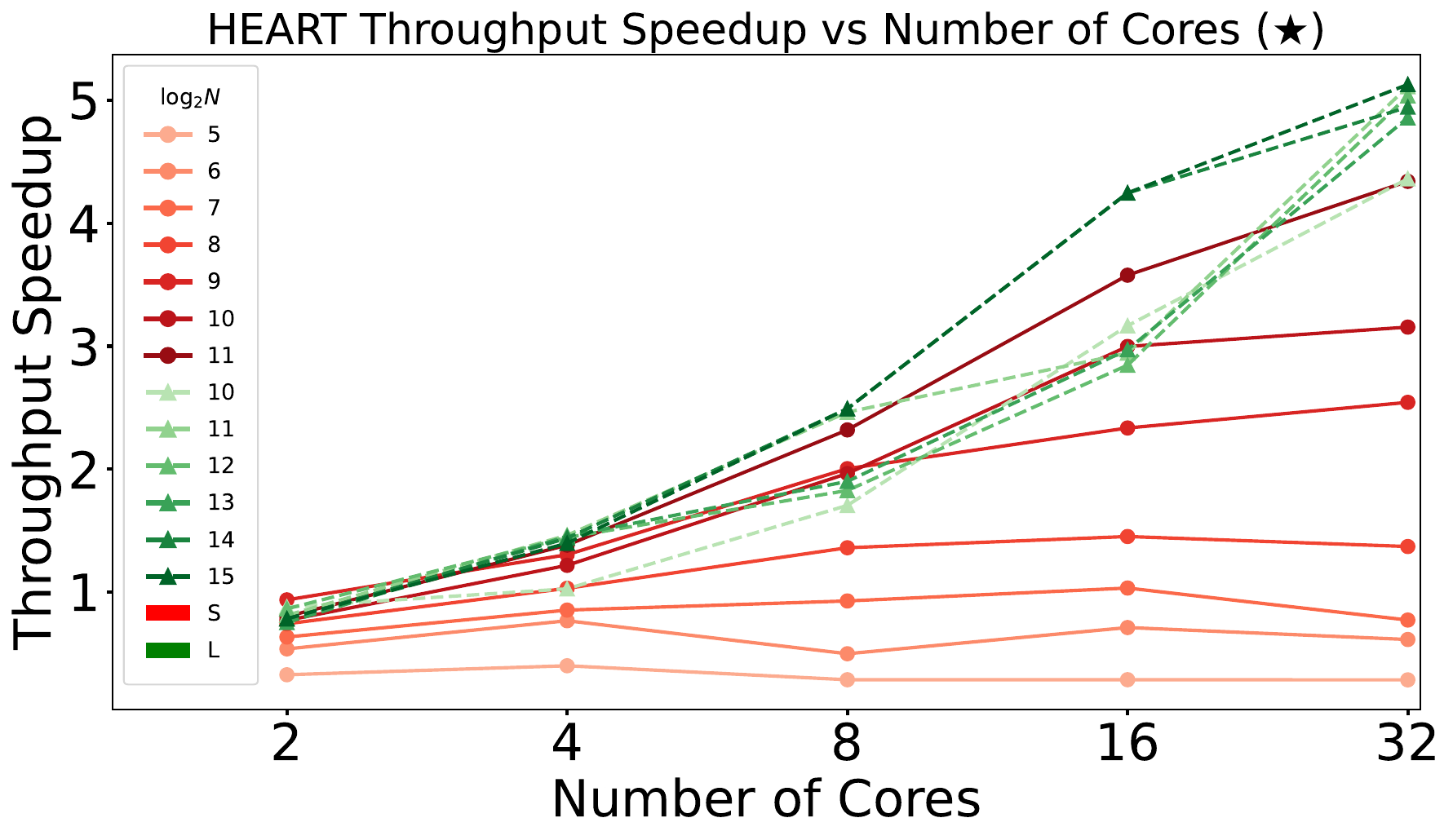}
        \subcaption{HEART (\ding{72})}
    \end{subfigure}%
    \hfill
    \begin{subfigure}[t]{0.48\textwidth}
        \centering
        \includegraphics[width=\linewidth]{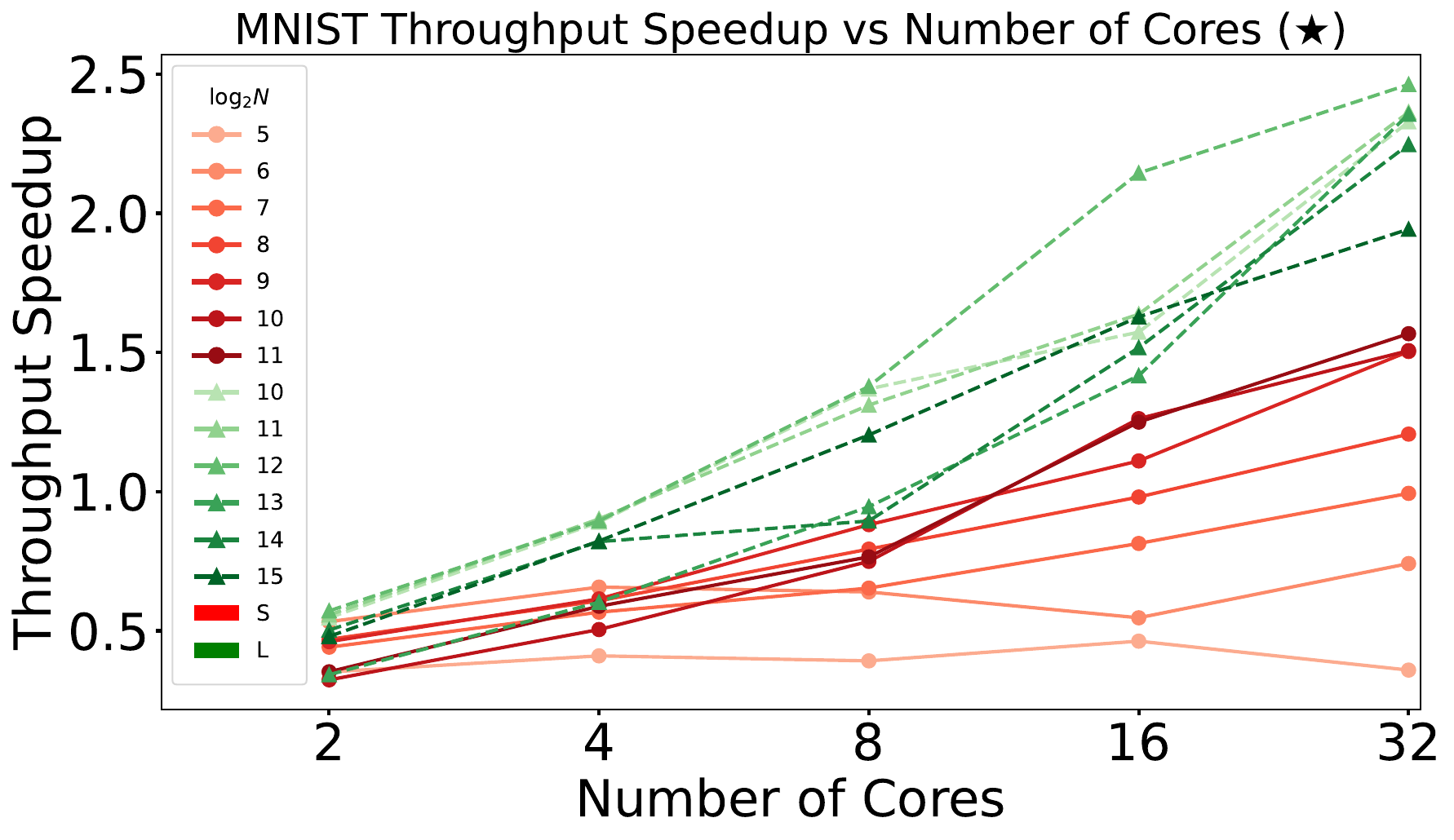}
        \subcaption{MNIST (\ding{72})}
    \end{subfigure}

    \vspace{0.75em}
    \begin{subfigure}[t]{0.48\textwidth}
        \centering
        \includegraphics[width=\linewidth]{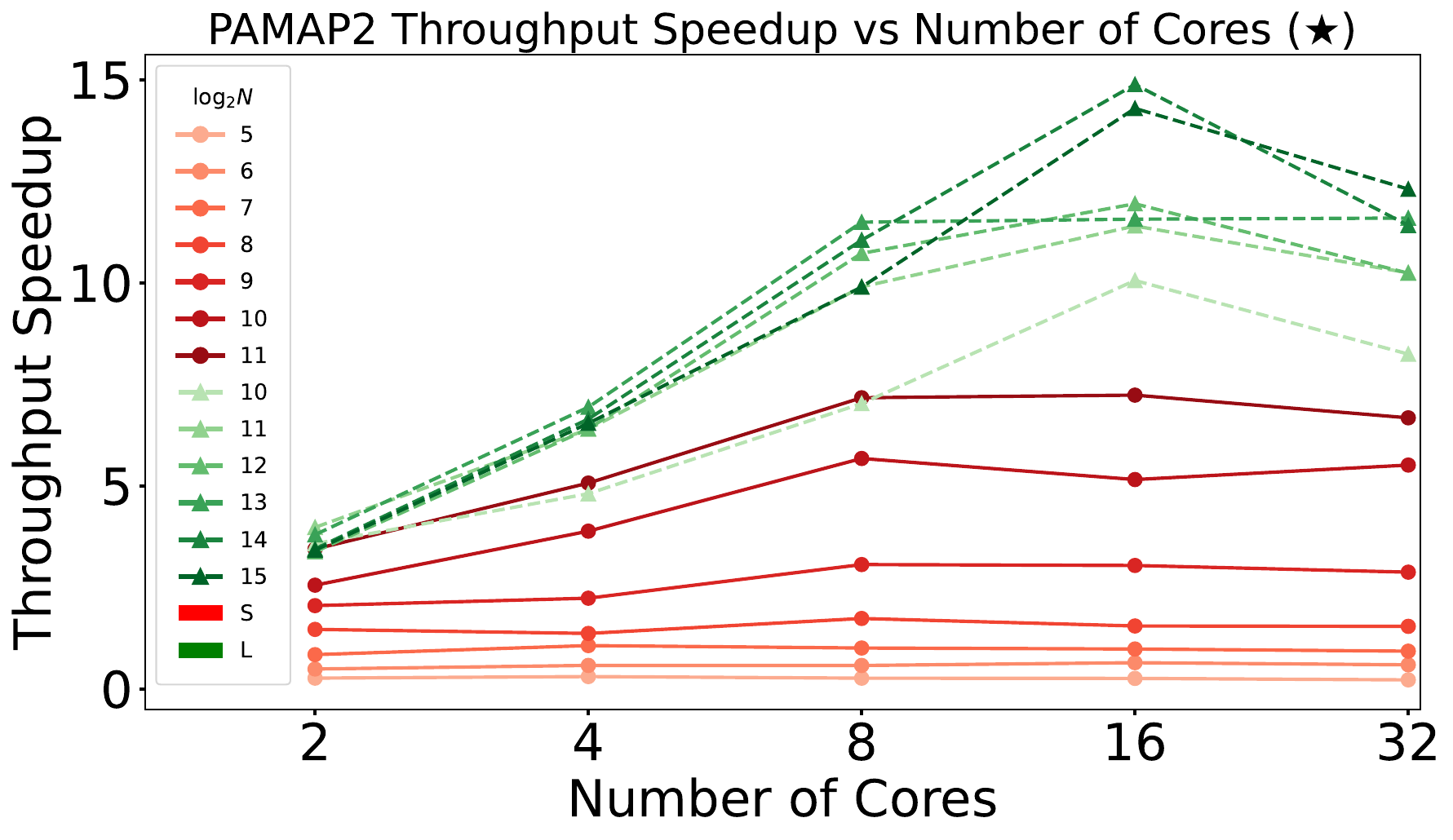}
        \subcaption{PAMAP2 (\ding{72})}
    \end{subfigure}%
    \hfill
    \begin{subfigure}[t]{0.48\textwidth}
        \centering
        \includegraphics[width=\linewidth]{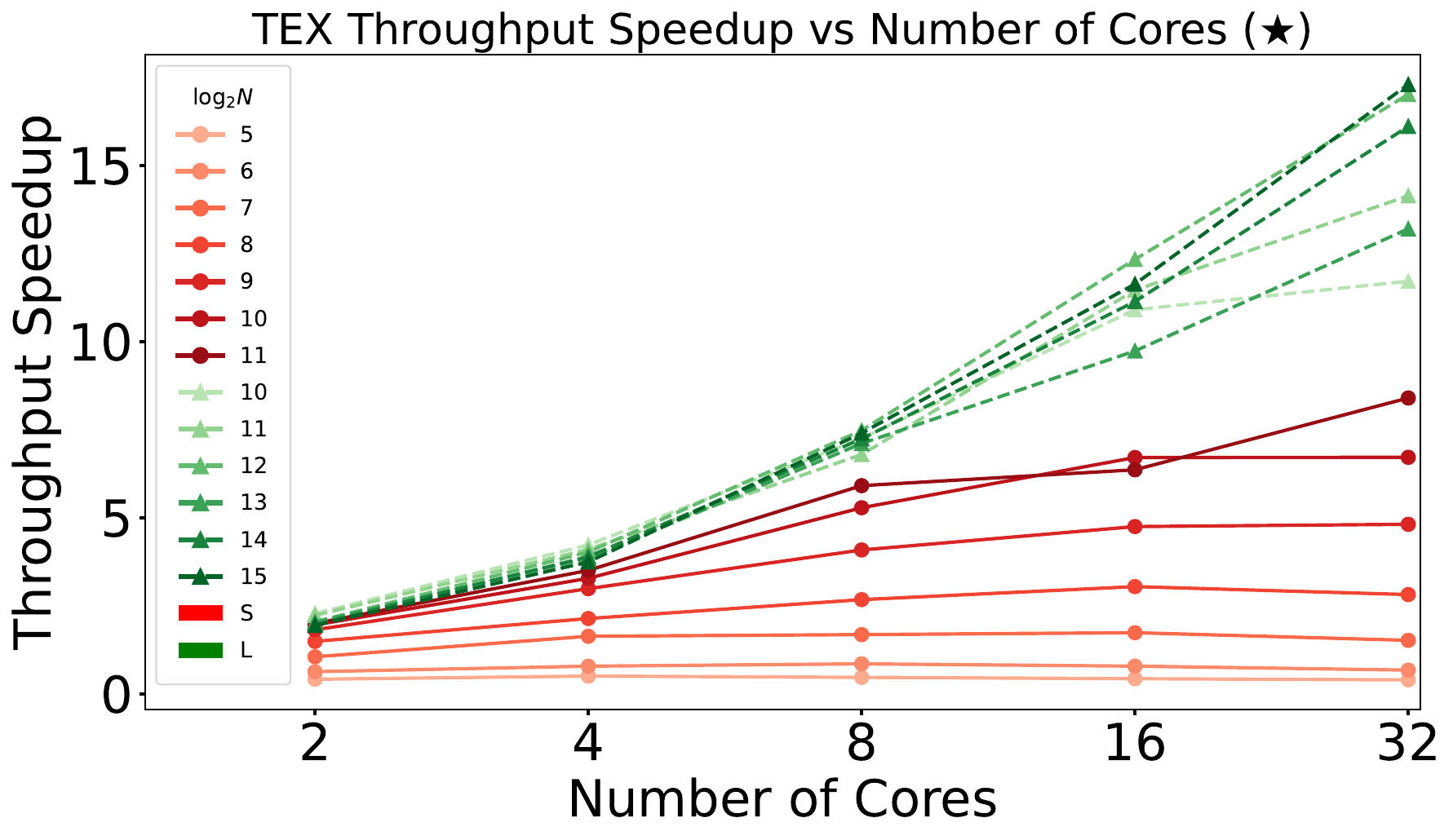}
        \subcaption{TEX (\ding{72})}
    \end{subfigure}

    % ========== INTEL Plots ==========
    \vspace{0.75em}
    \begin{subfigure}[t]{0.48\textwidth}
        \centering
        \includegraphics[width=\linewidth]{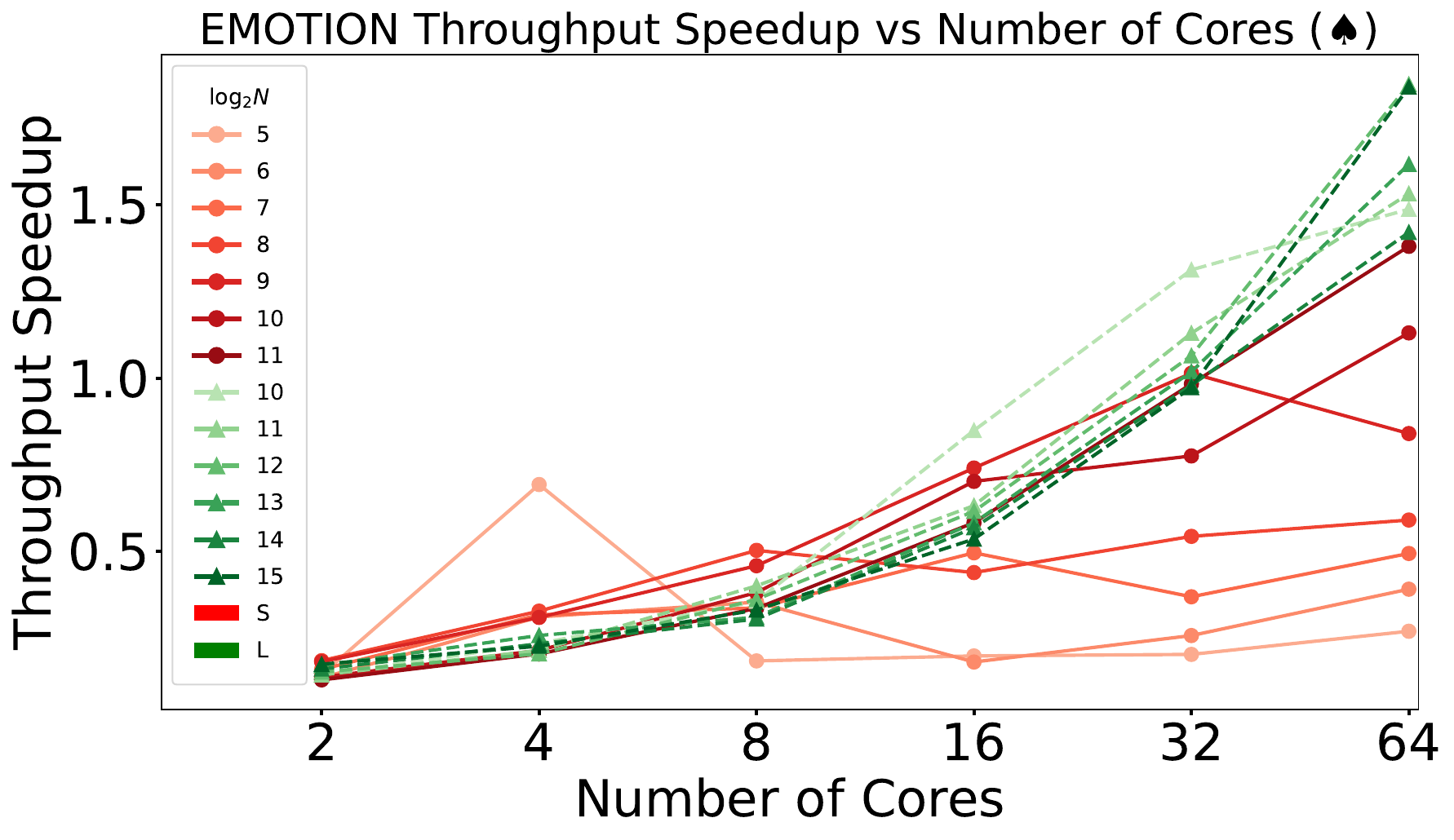}
        \subcaption{EMOTION (\ding{171})}
    \end{subfigure}%
    \hfill
    \begin{subfigure}[t]{0.48\textwidth}
        \centering
        \includegraphics[width=\linewidth]{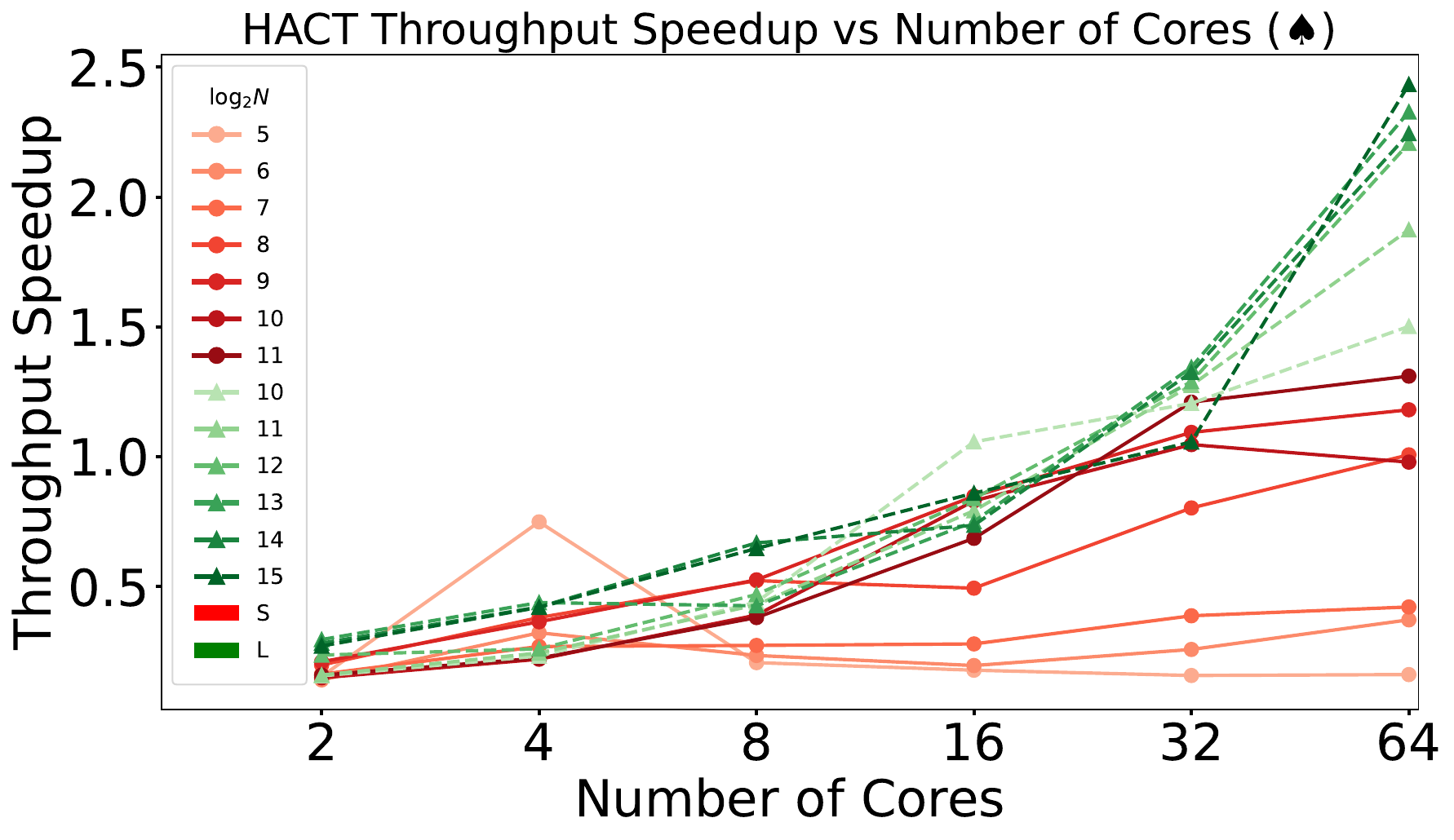}
        \subcaption{HACT (\ding{171})}
    \end{subfigure}

    \vspace{0.75em}
    \begin{subfigure}[t]{0.48\textwidth}
        \centering
        \includegraphics[width=\linewidth]{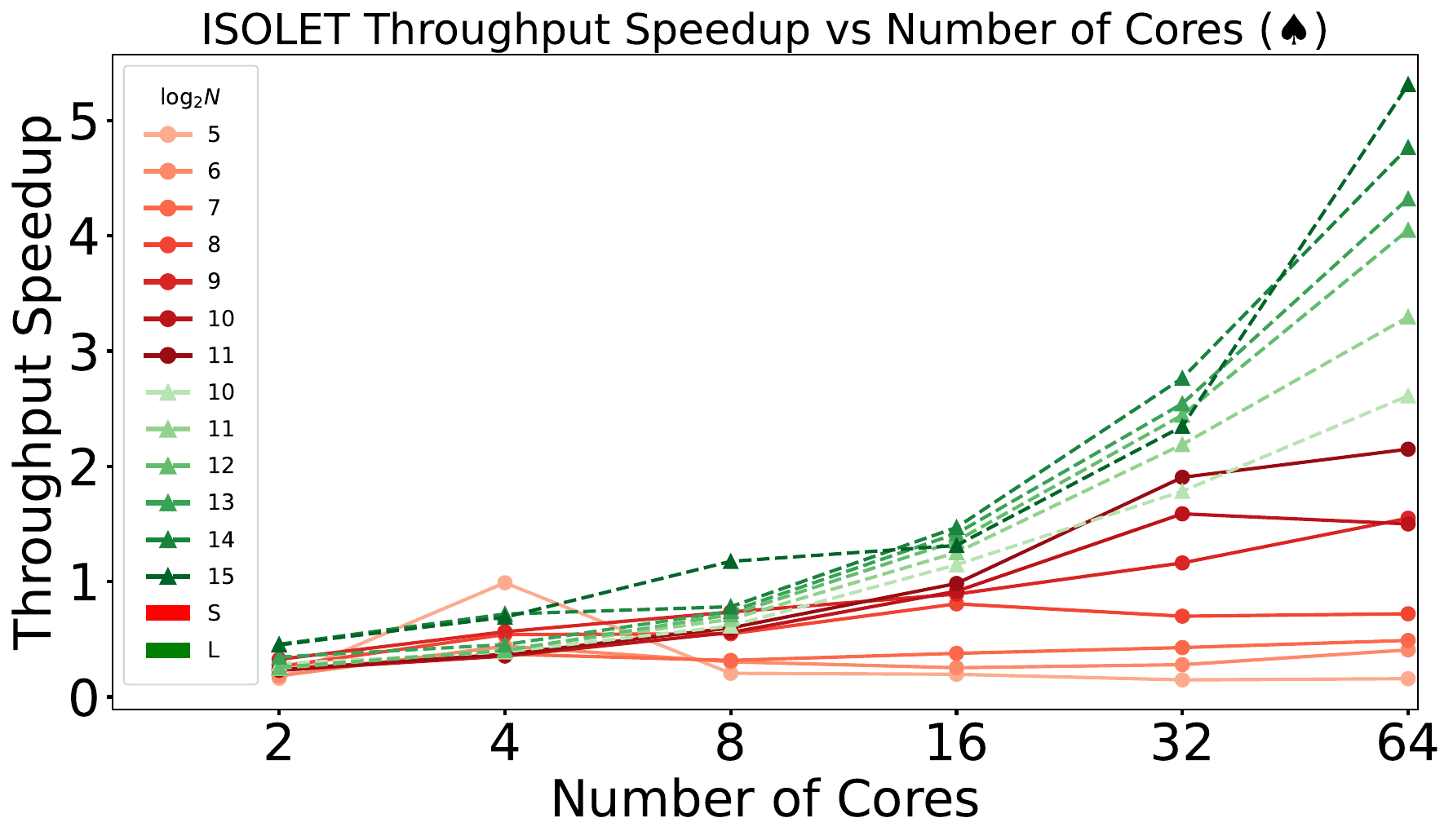}
        \subcaption{ISOLET (\ding{171})}
    \end{subfigure}%
    \hfill
    \begin{subfigure}[t]{0.48\textwidth}
        \centering
        \includegraphics[width=\linewidth]{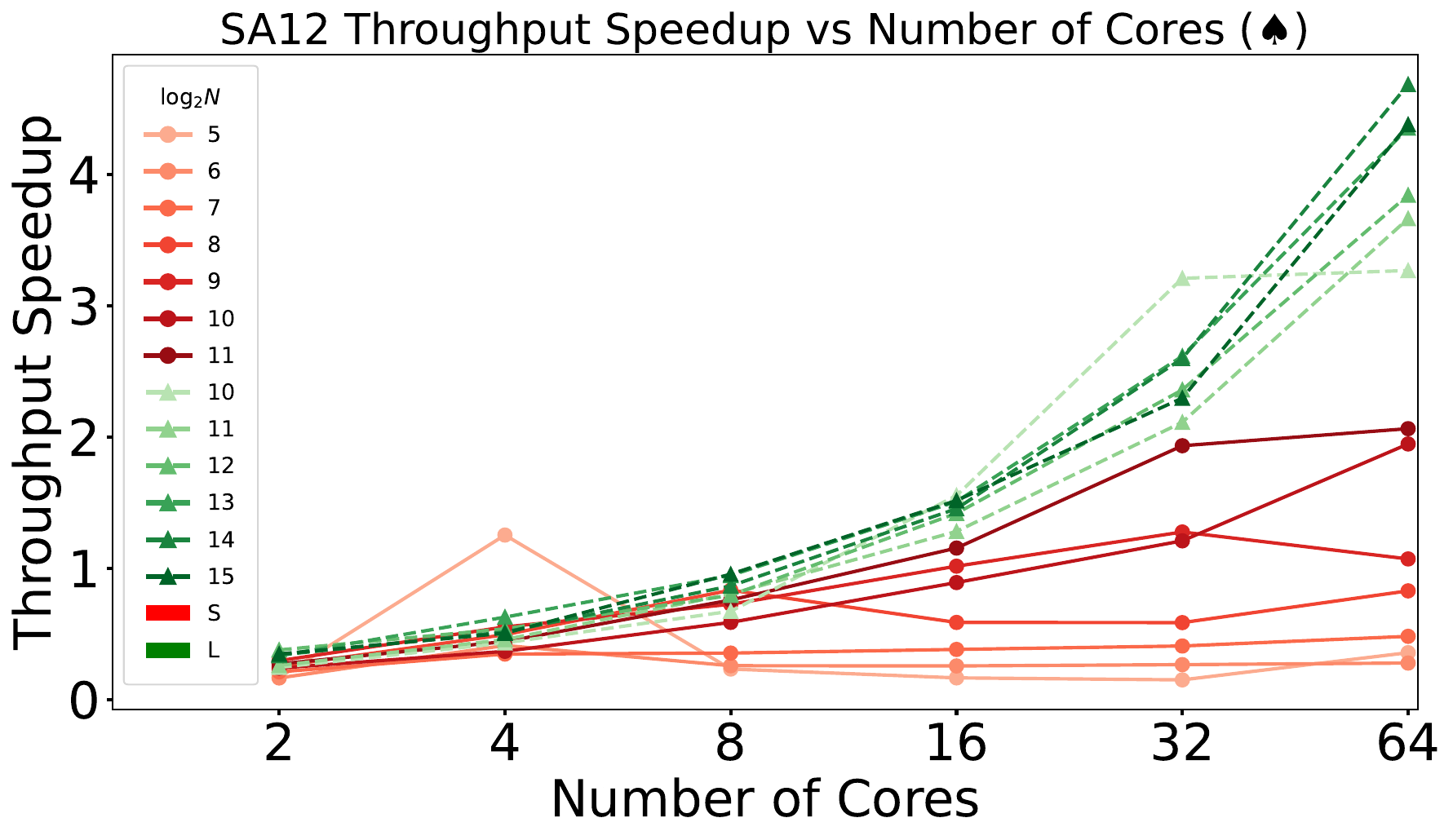}
        \subcaption{SA12 (\ding{171})}
    \end{subfigure}

    \caption{Speedup of ScalableHD over TorchHD for 8 datasets across AMD (\ding{72}) and Intel (\ding{171}) platforms. Note that $\mathrm{log}_2 N = 10, 11$ is shared between the two ScalableHD variants. Maximum throughput for ScalableHD across $(n, R)$ is compared against TorchHD baseline, both with $2T$ cores, yielding the speedup values used for plotting.}
    \label{fig:throughput_2x4}
\end{figure*}

For ScalableHD-S, we run inference with batch sizes $\mathrm{log}_2 N \in \{5, 6, 7, 8, 9, 10, 11\}$, whereas for ScalableHD-L we use $\mathrm{log}_2 N \in \{10, 11, 12, 13, 14, 15\}$. For each $(N, n, R)$ configuration, we evaluate ScalableHD inference performance across $2T \in \{ 2, 4, 8, \ldots, \gamma, 2\gamma \}$ total cores, assigning $T$ cores to workers in each pipeline stage. Note that SMT capability on both platforms (Section~\ref{subsec: plat}) is only utilized at $2T = 2\gamma$, where each stage receives $T$ physical threads (rather than cores). We use the standard TorchHD \cite{torchhd} library as our baseline. Built on top of PyTorch \cite{pytorch}, TorchHD is optimized for training and inference of HDC applications on both CPU and GPU platforms. To ensure fairness, we evaluate TorchHD over the same configurations of batch size and total cores, $(N, 2T)$, as used in ScalableHD.

We utilize throughput in samples per second as our performance metric \cite{torchhd, gpu-hdtorch, hetero-hpvm-hdc, cpu-hdcc-compiler, advance-on-cpu-ref}, defined for batch size $N$ as below,
\begin{equation}
\label{eq: thpt}
\text{throughput (samples/sec)} = \frac{1000}{N \cdot \text{latency (ms)}}
\end{equation}
The measured $\text{latency (ms)}$ refers to the end-to-end time in ms from when the input matrix $\mathbf{X} \in \mathbb{R}^{N \times F}$ is read from the external memory to when the predicted labels $\mathbf{y}_\text{pred} \in \mathbb{R}^{N}$ are written out to the external memory. We also utilize speedup as a performance metric, defined below,
\begin{equation}
\label{eq: speedup}
\text{speedup} = \frac{\text{throughput}_\text{ScalableHD}}{\text{throughput}_\text{baseline}}
\end{equation}
The speedup at batch size $N$ is computed as the ratio of the throughput achieved by ScalableHD to that of the TorchHD baseline, both measured at the same batch size $N$.

\subsection{Throughput Analysis}
\label{subsec: thru}

In this section, we discuss the performance of both variants of ScalableHD with respect to TorchHD. Fig. \ref{fig:throughput_2x4} shows the plots of throughput speedup (defined in Section \ref{subsec: setting}) versus number of cores. For each platform (defined in Section \ref{subsec: plat}), we select four of the eight datasets defined in Section \ref{subsec: dat} due to space constraints. Each plot shows the speedup for both, ScalableHD-S (red) and ScalableHD-L (green), across batch sizes. We measure throughput for both ScalableHD and TorchHD by averaging across $50$ runs. For a given $(N, 2T)$ pair, we compare the maximum throughput for ScalableHD, across the four combinations of $(n, R)$, versus the corresponding TorchHD throughput. Note that the number of cores for both ScalableHD and TorchHD is set as $2T$, ensuring fairness. The top two rows show the results for AMD (\ding{72}), whereas the bottom two rows show results for Intel (\ding{171}).

The plots clearly demonstrate ScalableHD's superior performance, when compared to TorchHD, across both multi-core CPU platforms. While TorchHD exhibits better performance than ScalableHD-S at small batch sizes ($N = 32, 64, 128$) and low core counts ($2T = 2, 4$), ScalableHD consistently outperforms TorchHD across all datasets as batch sizes and the number of utilized cores increase (see Table \ref{tab:throughput_summary}). This improvement stems from ScalableHD’s two-stage streaming pipeline and optimizations, which become increasingly effective at scale by mitigating data transfer and inter-core communication bottlenecks. For datasets such as HEART, ISOLET, and SA12, ScalableHD achieves speedups of up to $5\times$ over TorchHD, while for datasets like PAMAP2 and TEX, the speedup reaches up to $18\times$. Table \ref{tab:smt_effect} reports the impact of SMT on both ScalableHD and TorchHD across platforms. ScalableHD sees a slight performance gain when all logical CPUs are utilized, enabled by its NUMA-aware worker-to-core binding (except at $N=512$, which shows a slight drop). In contrast, TorchHD shows consistent degradation with SMT, due to thread contention. We next discuss ScalableHD's performance scalability.

\begin{table}[t]
\centering
\caption{Throughput and speedup (\textit{in parentheses}, w.r.t TorchHD) of ScalableHD-S and ScalableHD-L for selected datasets and batch sizes on AMD (\ding{72}) and Intel (\ding{171}) platforms.}
\label{tab:throughput_summary}
\renewcommand{\arraystretch}{1.2}    
\begin{adjustbox}{max width = 0.48\textwidth}
\begin{tabular}{|c|c|c|c|c|c|}
\hline
\textbf{Platform} & \textbf{Dataset} & \textbf{Variant} & $N$ & \textbf{Cores} ($2T$) & \textbf{Throughput}  ($\text{imgs}/\text{sec}$)\\ \hline

\multirow{24}{*}{Intel \ding{171}} 
  & \multirow{6}{*}{EMOTION} & \multirow{3}{*}{ScalableHD-S} & 256   & 64 & 2823.34 ($\times$0.59) \\ \cline{4-6}
  &                          &                                & 512   & 64 & 4257.63 ($\times$0.84) \\ \cline{4-6}
  &                          &                                & 1024  & 64 &  5689.86 ($\times$1.13)\\ \cline{3-6}
  &                          & \multirow{3}{*}{ScalableHD-L} & 2048  & 64 &   10408.43 ($\times$1.53)\\ \cline{4-6}
  &                          &                                & 8192  & 64 &  12689.93 ($\times$1.61)\\ \cline{4-6}
  &                          &                                & 32768 & 64 &  13702.31 ($\times$1.83)\\ \cline{2-6}
  & \multirow{6}{*}{HACT}    & \multirow{3}{*}{ScalableHD-S} & 256   & 64 &   4021.32 ($\times$1.00)\\ \cline{4-6}
  &                          &                                & 512   & 64 &   5224.56 ($\times$1.18)\\ \cline{4-6}
  &                          &                                & 1024  & 64 &   6733.37 ($\times$0.97)\\ \cline{3-6}
  &                          & \multirow{3}{*}{ScalableHD-L} & 2048  & 64 &   12693.24 ($\times$1.87)\\ \cline{4-6}
  &                          &                                & 8192  & 64 &   17944.85 ($\times$2.32)\\ \cline{4-6}
  &                          &                                & 32768 & 64 &   19824.96 ($\times$2.43)\\ \cline{2-6}

%\multirow{12}{*}{Intel \ding{171}} 
  & \multirow{6}{*}{ISOLET}  & \multirow{3}{*}{ScalableHD-S} & 256   & 64 &   4035.38 ($\times$0.72)\\ \cline{4-6}
  &                          &                                & 512   & 64 &   7077.22 ($\times$1.55)\\ \cline{4-6}
  &                          &                                & 1024  & 64 &   8309.56 ($\times$1.50)\\ \cline{3-6}
  &                          & \multirow{3}{*}{ScalableHD-L} & 2048  & 64 &   18550.38 ($\times$3.29)\\ \cline{4-6}
  &                          &                                & 8192  & 64 &   28952.36 ($\times$4.32)\\ \cline{4-6}
  &                          &                                & 32768 & 64 &  34464.53 ($\times$5.31)\\ \cline{2-6}
  & \multirow{6}{*}{SA12}    & \multirow{3}{*}{ScalableHD-S} & 256   & 64 & 4032.32  ($\times$0.82)\\ \cline{4-6}
  &                          &                                & 512   & 64 &  5979.69  ($\times$1.07)\\ \cline{4-6}
  &                          &                                & 1024  & 64 &  11253.47 ($\times$1.94)\\ \cline{3-6}
  &                          & \multirow{3}{*}{ScalableHD-L} & 2048  & 64 &   24651.85 ($\times$3.66)\\ \cline{4-6}
  &                          &                                & 8192  & 64 &   31664.53 ($\times$4.35)\\ \cline{4-6}
  &                          &                                & 32768 & 64 &  34278.58 ($\times$4.37)\\ \hline

\multirow{24}{*}{AMD \ding{72}}   
  & \multirow{6}{*}{HEART}   & \multirow{3}{*}{ScalableHD-S} & 256   & 32 &   15689.84 ($\times$1.31)\\ \cline{4-6}
  &                          &                                & 512   & 32 &   29186.47 ($\times$2.72)\\ \cline{4-6}
  &                          &                                & 1024  & 32 &   38159.73 ($\times$3.07)\\ \cline{3-6}
  &                          & \multirow{3}{*}{ScalableHD-L} & 2048  & 32 &  61799.04 ($\times$5.06)\\ \cline{4-6}
  &                          &                                & 8192  & 32 &   61439.19 ($\times$4.75)\\\cline{4-6}
  &                          &                                & 32768 & 32 &   63220.32 ($\times$4.92)\\ \cline{2-6}
  & \multirow{6}{*}{MNIST}   & \multirow{3}{*}{ScalableHD-S} & 256   & 32 &  11352.28 ($\times$1.17)\\\cline{4-6}
  &                          &                                & 512   & 32 &  14379.04 ($\times$1.70)\\\cline{4-6}
  &                          &                                & 1024  & 32 &  15636.94 ($\times$1.50)\\ \cline{3-6}
  &                          & \multirow{3}{*}{ScalableHD-L} & 2048  & 32 &  24776.12 ($\times$2.36)\\\cline{4-6}
  &                          &                                & 8192  & 32 &  24747.44 ($\times$2.28)\\\cline{4-6}
  &                          &                                & 32768 & 32 &  20540.04 ($\times$1.93)\\ \cline{2-6}

%\multirow{12}{*}{AMD \ding{72}}   
  & \multirow{6}{*}{PAMAP2}  & \multirow{3}{*}{ScalableHD-S} & 256   & 32 &  17417.51 ($\times$1.46)\\\cline{4-6}
  &                          &                                & 512   & 32 &  33619.08 ($\times$3.03)\\\cline{4-6}
  &                          &                                & 1024  & 32 &   65261.58 ($\times$5.13)\\ \cline{3-6}
  &                          & \multirow{3}{*}{ScalableHD-L} & 2048  & 32 &   130513.60 ($\times$10.16)\\\cline{4-6}
  &                          &                                & 8192  & 32 &   151928.76 ($\times$11.50)\\\cline{4-6}
  &                          &                                & 32768 & 32 &    157977.37 ($\times$12.16)\\ \cline{2-6}
  & \multirow{6}{*}{TEX}     & \multirow{3}{*}{ScalableHD-S} & 256   & 32 &   18805.00 ($\times$2.71)\\\cline{4-6}
  &                          &                                & 512   & 32 &   32190.11 ($\times$4.35)\\\cline{4-6}
  &                          &                                & 1024  & 32 &   45732.73 ($\times$6.00)\\ \cline{3-6}
  &                          & \multirow{3}{*}{ScalableHD-L} & 2048  & 32 &   105690.65 ($\times$13.79)\\\cline{4-6}
  &                          &                                & 8192  & 32 &   101170.27 ($\times$12.89)\\\cline{4-6}
  &                          &                                & 32768 & 32 &   129213.04 ($\times$17.64)\\ \hline

\end{tabular}
\end{adjustbox}
\end{table}

\begin{table}[t]
\centering
\caption{Impact of SMT on throughput. Percentage change in throughput is measured when moving from using all physical cores ($2T = \gamma$) to all SMT threads ($2T = 2\gamma$), for both ScalableHD and TorchHD.}
\label{tab:smt_effect}
\begin{adjustbox}{max width = 0.48\textwidth}
\renewcommand{\arraystretch}{1.2}
\begin{tabular}{|c|c|c|c|c|c|}
\hline
\textbf{Platform} & \textbf{Dataset} & \textbf{Variant} & $N$ & \textbf{ScalableHD (\%$\Delta$)} & \textbf{TorchHD (\%$\Delta$)} \\ \hline

\multirow{4}{*}{Intel \ding{171}} & \multirow{4}{*}{HACT} 
& \multirow{2}{*}{ScalableHD-S} & 512   & $-3.03\%$ & $-7.05\%$ \\ \cline{4-6}
&            &                   & 1024  & $+3.63\%$ & $-21.12\%$ \\ \cline{3-6}
&                           & \multirow{2}{*}{ScalableHD-L} & 8192  & $+5.14\%$ & $-3.52\%$  \\ \cline{4-6}
&                           &  & 32768 & $+6.59\%$ & $-7.2\%$ \\ \hline

\multirow{4}{*}{AMD \ding{72}} & \multirow{4}{*}{MNIST} 
& \multirow{2}{*}{ScalableHD-S} & 512   & $-3.98\%$ & $-6.97\%$ \\ \cline{4-6}
&                           &  & 1024  & $+3.67\%$ & $-14.5\%$ \\ \cline{3-6}
&                           & \multirow{2}{*}{ScalableHD-L} & 8192  & $+20.29\%$ & $-1.8\%$ \\ \cline{4-6}
&                           &  & 32768 & $+46.13\%$ & $+5.48\%$ \\ \hline

\end{tabular}
\end{adjustbox}
\end{table}

\subsection{Scalability Analysis}
\label{subsec: scal}

\begin{figure*}
    \centering
    % ========== AMD Scalability Plots ==========
    \begin{subfigure}[t]{0.48\textwidth}
        \centering
        \includegraphics[width=\linewidth]{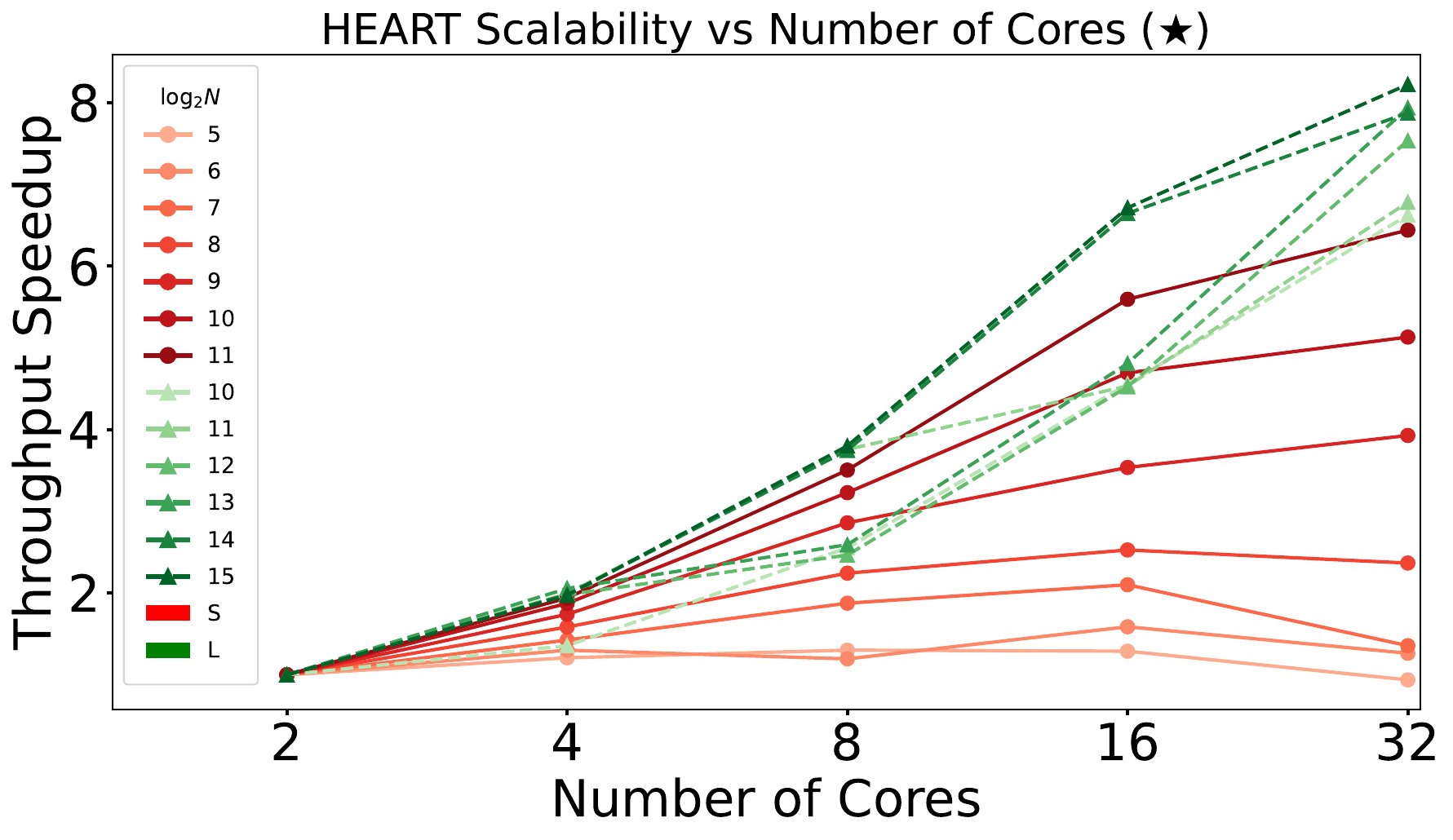}
        \subcaption{HEART (\ding{72})}
    \end{subfigure}%
    \hfill
    \begin{subfigure}[t]{0.48\textwidth}
        \centering
        \includegraphics[width=\linewidth]{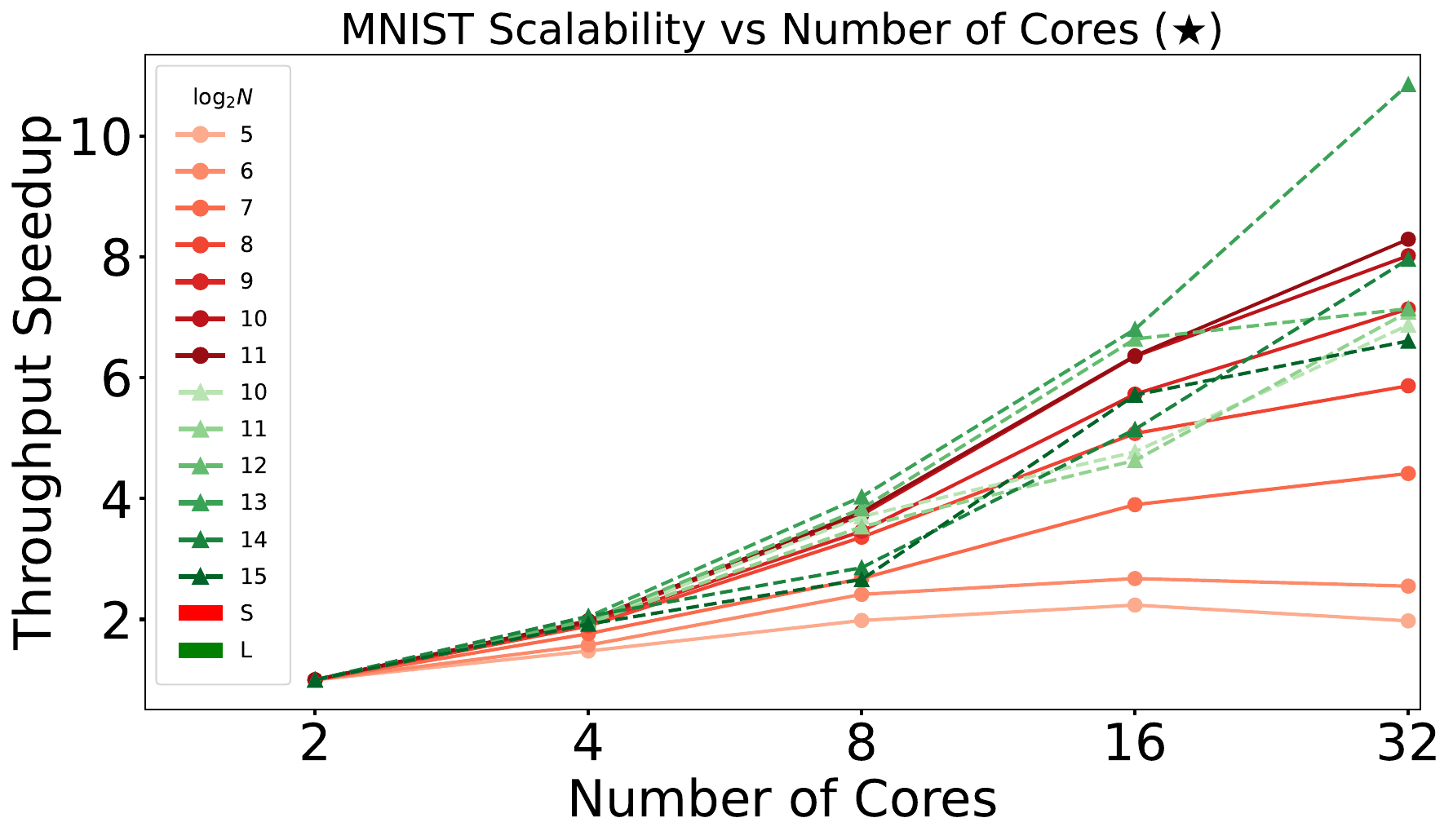}
        \subcaption{MNIST (\ding{72})}
    \end{subfigure}

    \vspace{0.75em}
    \begin{subfigure}[t]{0.48\textwidth}
        \centering
        \includegraphics[width=\linewidth]{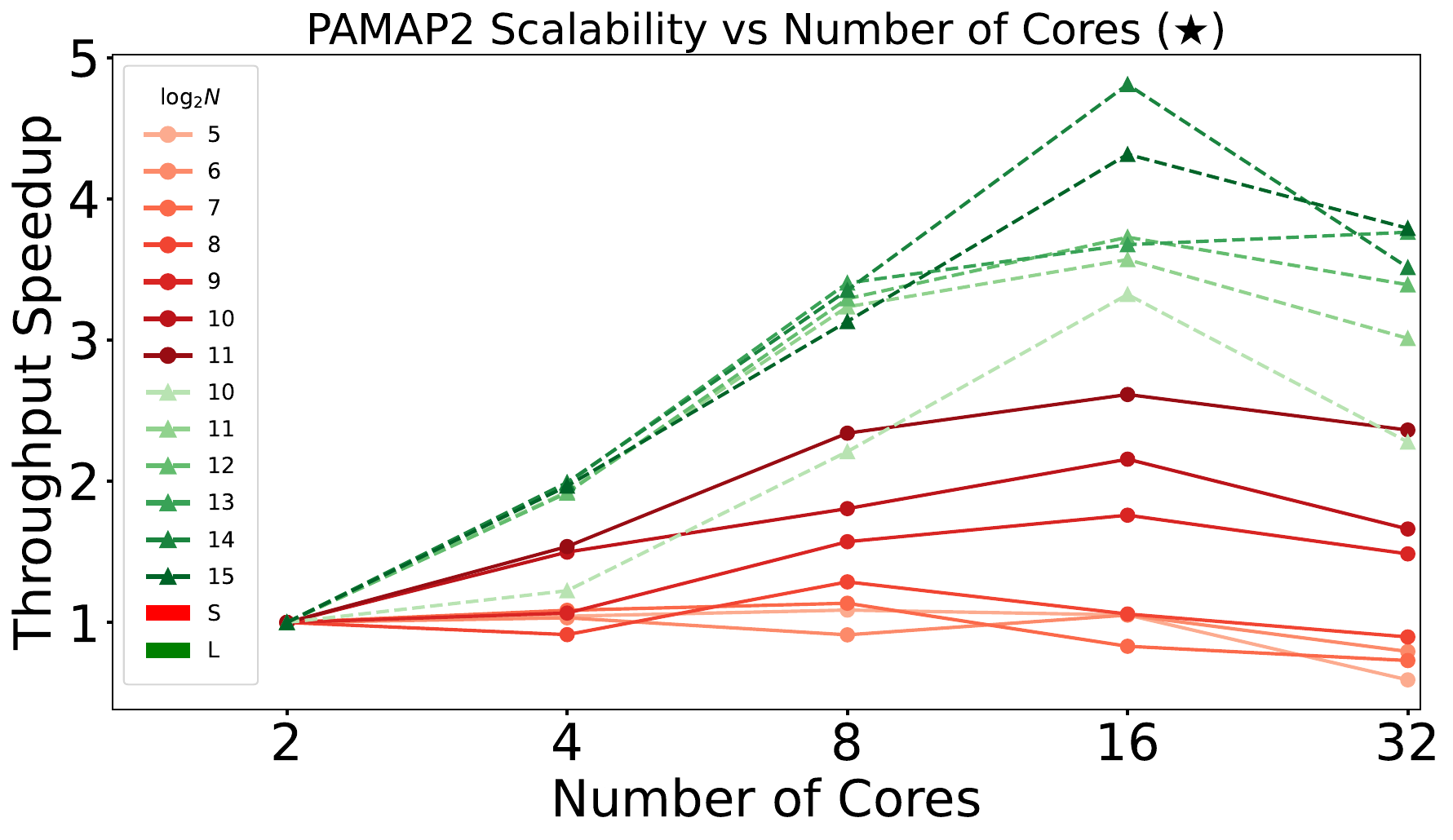}
        \subcaption{PAMAP2 (\ding{72})}
    \end{subfigure}%
    \hfill
    \begin{subfigure}[t]{0.48\textwidth}
        \centering
        \includegraphics[width=\linewidth]{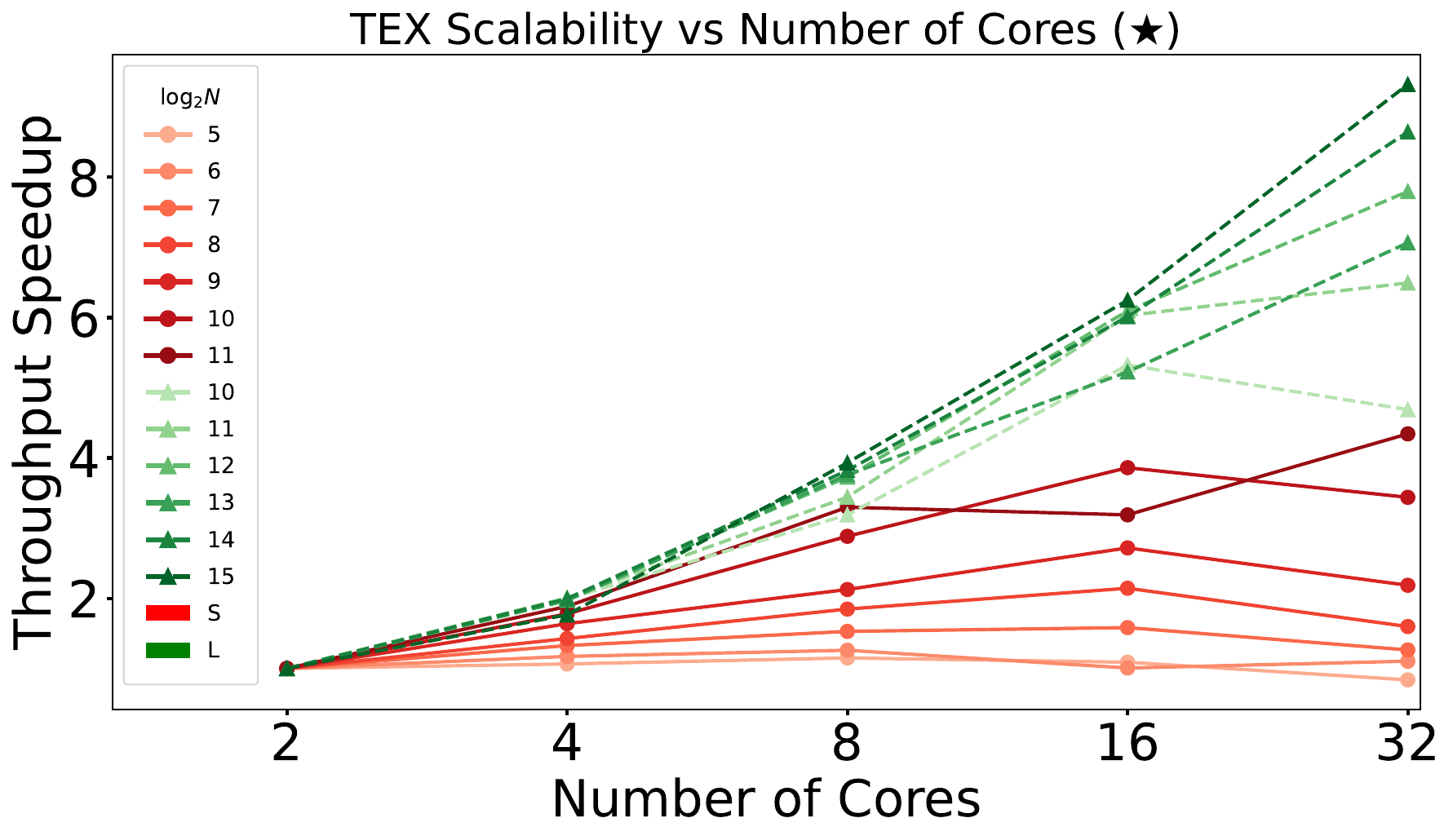}
        \subcaption{TEX (\ding{72})}
    \end{subfigure}

    % ========== INTEL Scalability Plots ==========
    \vspace{0.75em}
    \begin{subfigure}[t]{0.48\textwidth}
        \centering
        \includegraphics[width=\linewidth]{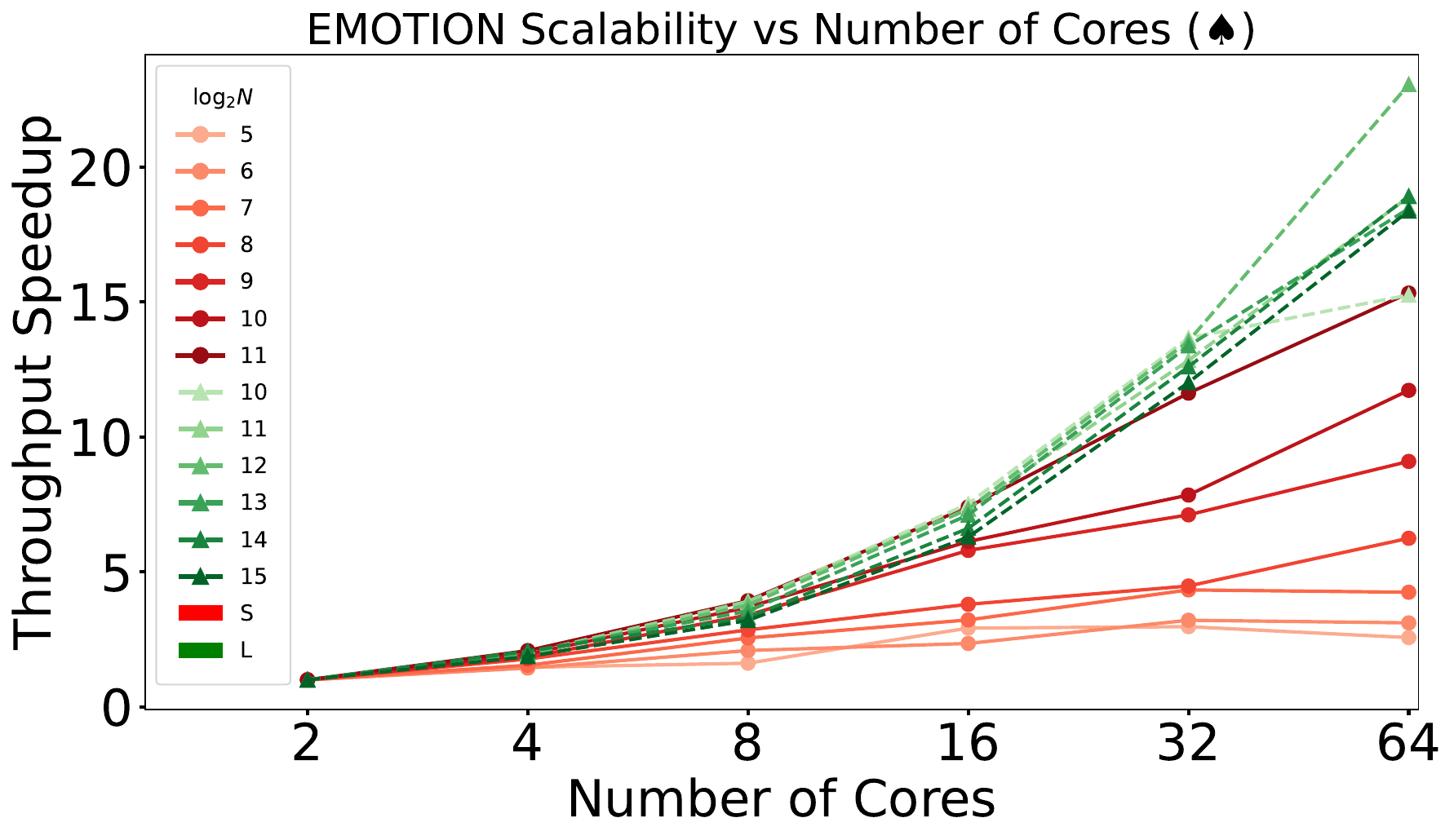}
        \subcaption{EMOTION (\ding{171})}
    \end{subfigure}%
    \hfill
    \begin{subfigure}[t]{0.48\textwidth}
        \centering
        \includegraphics[width=\linewidth]{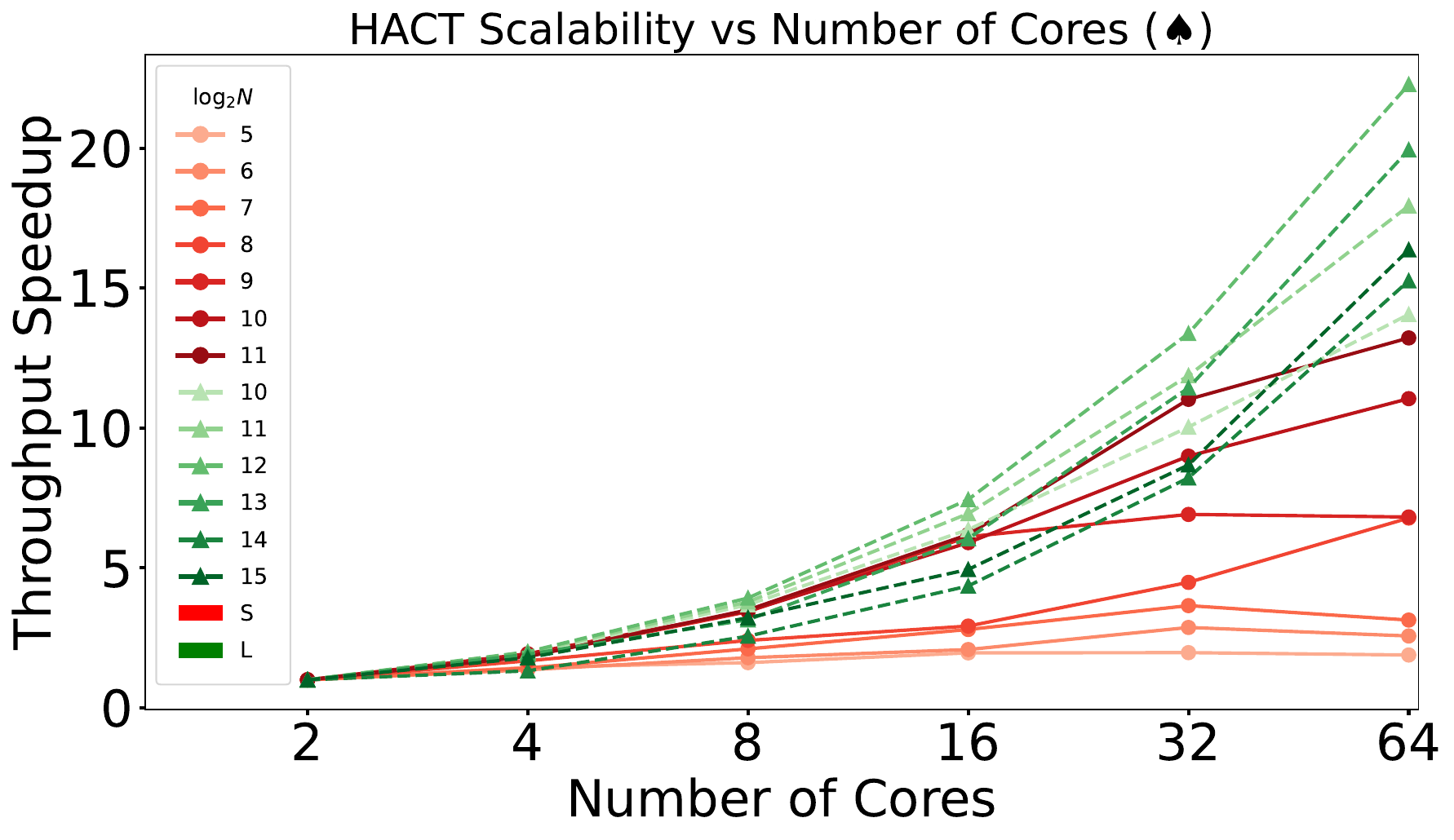}
        \subcaption{HACT (\ding{171})}
    \end{subfigure}

    \vspace{0.75em}
    \begin{subfigure}[t]{0.48\textwidth}
        \centering
        \includegraphics[width=\linewidth]{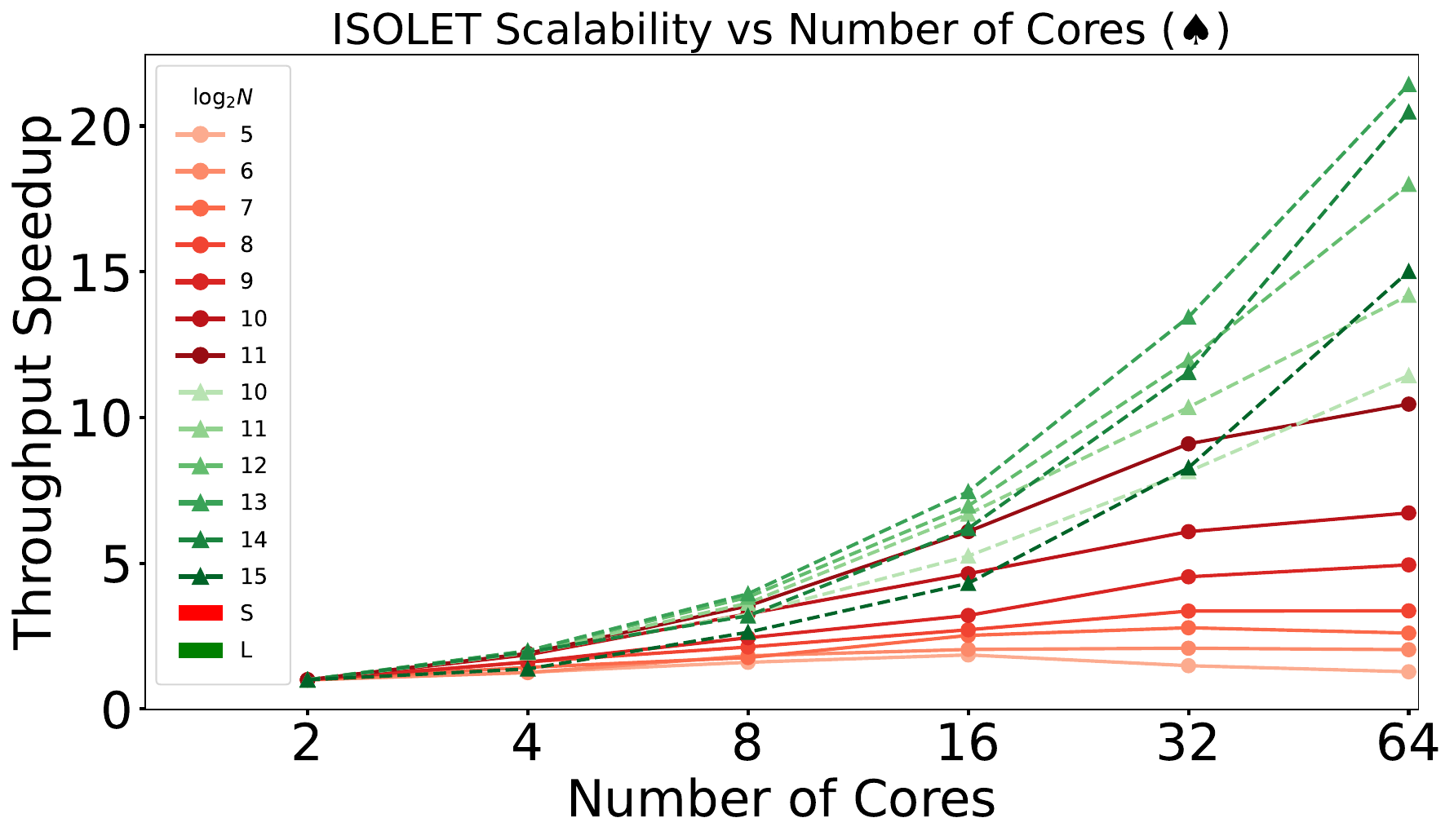}
        \subcaption{ISOLET (\ding{171})}
    \end{subfigure}%
    \hfill
    \begin{subfigure}[t]{0.48\textwidth}
        \centering
        \includegraphics[width=\linewidth]{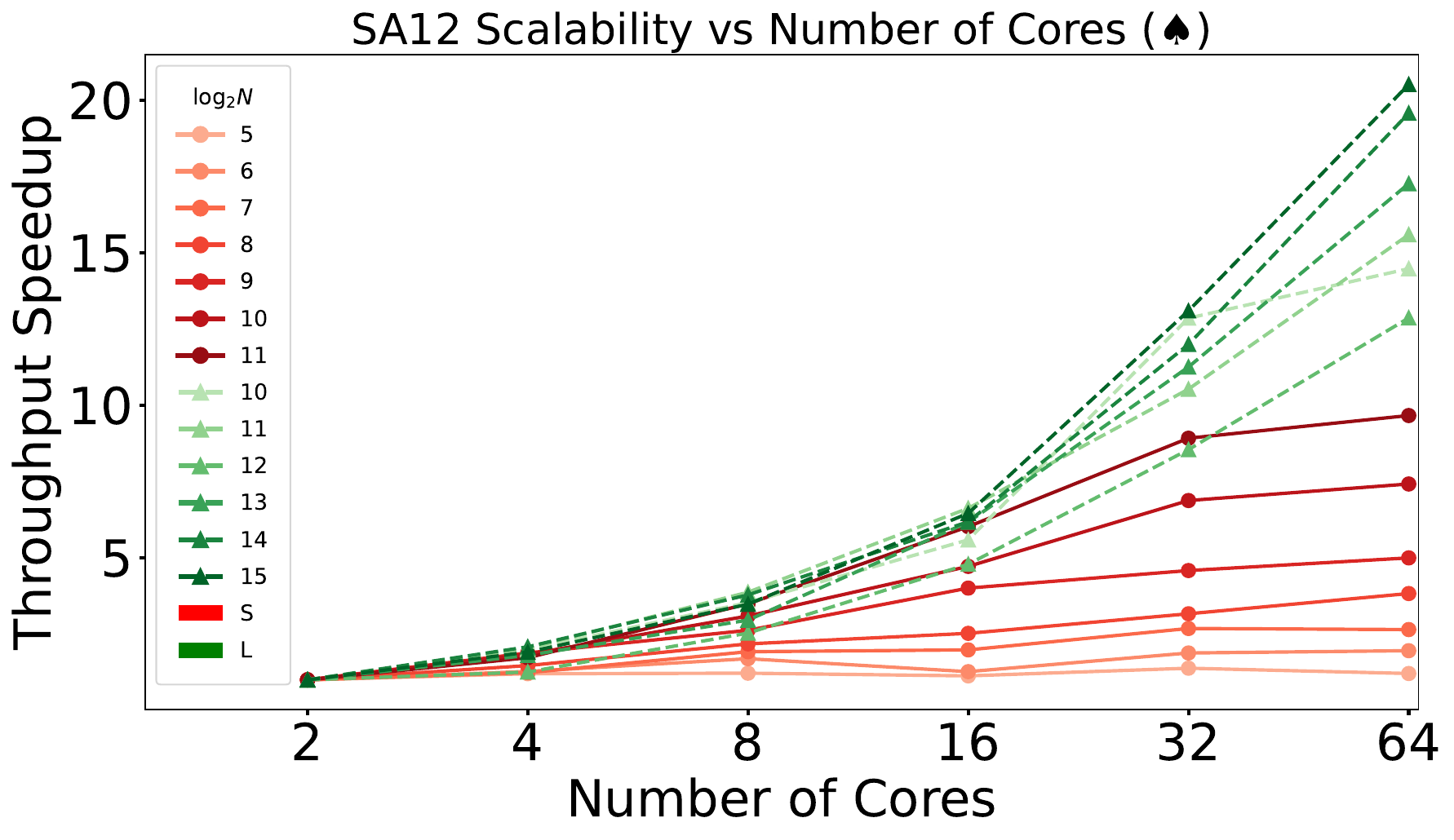}
        \subcaption{SA12 (\ding{171})}
    \end{subfigure}

    \caption{Scalability of ScalableHD across number of cores for 8 datasets on AMD (\ding{72}) and Intel (\ding{171}) platforms. Each plot shows throughput speedup (relative to 2-core execution) for both ScalableHD variants across multiple batch sizes. The x-axis is log-scaled; tile size $n = 32$ and chunk count $R = 8$ are fixed for all runs.}
    \label{fig:scalability_2x4}
\end{figure*}

Fig. \ref{fig:scalability_2x4} demonstrates the scalability performance for both variants of ScalableHD. We select four datasets per platform, as before, and plot the throughput speedup versus total number of cores. Note that for each curve, the throughput at $2T=2$ cores is used as baseline to measure the speedup. We fix $n=32$ and $R=8$ for these plots. As is evident from Fig. \ref{fig:scalability_2x4}, ScalableHD, for both variants, demonstrates near linear performance scaling with increasing core count. Note that the x-axis for all the plots is $\log$ scale, thus the curves are roughly $\log$-linear. This near $\log$-linear scaling of ScalableHD is enabled by the two-stage pipeline, where the large intermediate encoded matrix $\mathbf{H}$ is streamed in portions across the workers (cores) of both stages. Memory tiling ensures efficient data access, while optimized worker-to-core binding reduces inter-core communication overhead. Further note that the scaling trend holds across a wide range of batch sizes, demonstrating that ScalableHD maintains efficient utilization of increasing core counts for both S and L variants.

\subsection{Ablation Study}
\label{subsec: abl}
\begin{figure}
    \centering
    \includegraphics[width=\linewidth]{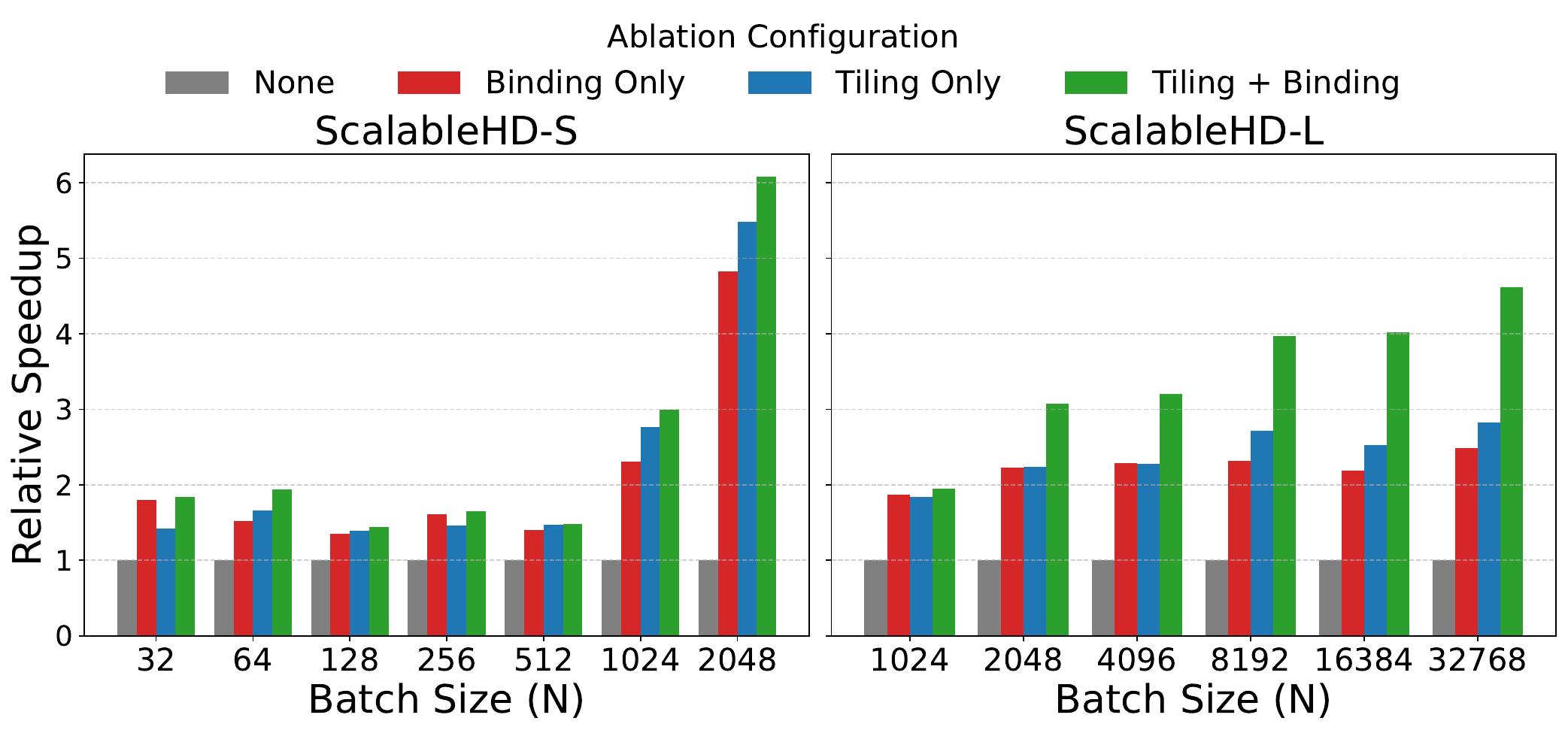}
    \caption{Ablation study analyzing the impact of optimizations}
    \label{fig: ablation}
\end{figure}

Figure~\ref{fig: ablation} presents an ablation analysis of ScalableHD’s core optimizations: memory tiling and NUMA-aware worker-to-core binding. We conduct experiments on the Intel (\ding{171}) platform using the MNIST dataset with $2T = 64$ total threads, and plot the relative throughput speedup for each configuration. The baseline corresponds to the unoptimized version (no tiling, no NUMA-aware binding) and is normalized to $1\times$. We observe that both optimizations contribute significantly to performance, with memory tiling providing greater standalone gains than binding, particularly at larger batch sizes. For example, in ScalableHD-L, tiling alone yields up to $2.8\times$ speedup, while binding alone yields up to $2.5\times$. The combined configuration consistently outperforms the individual ones, achieving up to $6.1\times$ speedup for ScalableHD-S and $4.6\times$ for ScalableHD-L. At smaller batch sizes, gains are more modest, as the compute workload is insufficient to fully leverage the benefits of tiling and inter-core communication optimization. However, as batch size increases, the impact of both optimizations becomes pronounced, resulting in substantial throughput improvements.

\section{Related Work}
\label{sec: related}

HDC has recently emerged as a lightweight, parallelizable alternative to compute-intensive Deep Neural Networks (DNNs), offering robustness to noise, low memory footprint, and suitability for real-time inference on resource-constrained edge platforms \cite{kanerva, survey-two-parter-part-I, survey-hdc-stoch-framework, survey-hdc-edge-intel-progress}. Recent efforts have focused on improving HDC model accuracy through learned encodings and adaptive training \cite{trainableHD, static-encoding-6-genie-hd, static-encoding-7-efficient-human-activity-recognition-hdc-imani, static-encoding-9-hdc-framework-image-descriptors-cvpr}. Substantial work has also focused on optimizing HDC inference across a wide range of hardware platforms\textemdash including FPGAs, GPUs, PIM/CIM architectures, and CPUs\textemdash to meet the performance demands of real-world applications.

\vspace{2pt}
\noindent \textbf{FPGA} Many early research efforts in HDC acceleration target FPGA platforms for energy-efficient and low-latency inference \cite{fpga-e3-hdc, fpga-edge-ai-acc, fpga-general-hdc-edge, fpga-hdc-hardware-acc-cvpr, fpga-imani-revisiting-hdc-fpga, fpga-onlinehd, fpga-scalable-interpretable}. These accelerators exploit the fine-grained parallelism inherent to HDC operations, implementing them as pipelined dataflow architectures or parallelized compute kernels \cite{imani-brain-HDC}. Recent FPGA-based designs also emphasize interpretability and scalability for edge applications, as demonstrated in \cite{fpga-scalable-interpretable, fpga-onlinehd}.

\vspace{2pt}
\noindent \textbf{PIM/CIM} Several works also explore emerging memory-centric architectures like Processor-In-Memory (PIM) and Compute-In-Memory (CIM) to mitigate memory bottlenecks associated with high-dimensional HVs in HDC inference \cite{pim-biohd-imani, pim-dual-accl-cluster-pim, pim-hdnn-rosing}. These platforms offer low-latency, high-bandwidth data movement but require specialized hardware, limiting their general applicability.

\vspace{2pt}
\noindent \textbf{GPU} More recently, HDC inference has been accelerated on GPUs. Libraries such as OpenHD, HDTorch \cite{gpu-hdtorch, gpu-openhd}, and TorchHD \cite{torchhd} provide GPU-optimized kernels for core HDC operations, while XcelHD \cite{gpu-xcel-hd-efficient-gpu} explores throughput-driven optimizations. Although these works achieve notable performance, they often lack generality across diverse tasks and applications, and require access to specialized GPU hardware for deployment.

\vspace{2pt}
\noindent \textbf{Multi-Platform} Few frameworks support multiple hardware backends \cite{multi-comphd, multi-mani-hd-imani, multi-multiarch-hardware-acc, hetero-hpvm-hdc}, including deployment on heterogeneous platforms. However, these works primarily exploit coarse-grained task-level parallelism and do not incorporate fine-grained, platform-specific optimizations, limiting their scalability across applications and hardware configurations.

\vspace{2pt}
\noindent \textbf{CPU} Despite the ubiquity of multi-core CPUs, only a handful of works address CPU-based acceleration. HDCC \cite{cpu-hdcc-compiler} introduces a compiler for portable C-style HDC deployment, but does not target throughput or scalability. Prior CPU implementations parallelize HDC inference naively using multi-threading, resulting in poor throughput and limited scalability due to unoptimized memory access and high inter-core communication overheads \cite{advance-on-cpu-ref}.

\vspace{2pt}
\noindent \textbf{Need for General-Purpose \& Scalable Inference} Real-world deployments of HDC—such as in human activity recognition, bio-signal classification, or real-time emotion detection—require robust high-throughput and scalable inference pipelines on general-purpose CPUs \cite{static-encoding-2-seizure-det, static-encoding-7-efficient-human-activity-recognition-hdc-imani, pim-biohd-imani, advance-on-cpu-ref, hetero-hpvm-hdc, cpu-hdcc-compiler}. ScalableHD directly addresses this gap by introducing a two-stage pipelined execution strategy and performance-aware optimizations such as memory tiling and NUMA-aware binding—enabling efficient, high-throughput HDC inference on widely available multi-core CPU platforms.
\section{Conclusion}
\label{sec: concl}
We presented ScalableHD, a two-stage pipelined framework for efficient and scalable HDC inference on general-purpose multi-core CPUs. By combining memory tiling, NUMA-aware core binding, and streaming inter-stage execution, ScalableHD delivers up to $10\times$ throughput gains over state-of-the-art TorchHD CPU baseline, without sacrificing accuracy. Our design supports both small as well as large batches, scaling near-linearly with increasing core counts. Future work includes extending ScalableHD with dynamic load balancing, supporting heterogeneous CPU–GPU execution, and exploring FPGA-based acceleration for end-to-end HDC inference pipelines.

\section*{Acknowledgement}
This work is supported by the National Science Foundation (NSF) under grant OAC-2209563.
%\section*{Acknowledgment}

%%% remove after completing paper
%\clearpage
\bibliographystyle{IEEEtran}
\bibliography{IEEEabrv, ref}

\begin{thebibliography}{10}
\providecommand{\url}[1]{#1}
\csname url@samestyle\endcsname
\providecommand{\newblock}{\relax}
\providecommand{\bibinfo}[2]{#2}
\providecommand{\BIBentrySTDinterwordspacing}{\spaceskip=0pt\relax}
\providecommand{\BIBentryALTinterwordstretchfactor}{4}
\providecommand{\BIBentryALTinterwordspacing}{\spaceskip=\fontdimen2\font plus
\BIBentryALTinterwordstretchfactor\fontdimen3\font minus \fontdimen4\font\relax}
\providecommand{\BIBforeignlanguage}[2]{{%
\expandafter\ifx\csname l@#1\endcsname\relax
\typeout{** WARNING: IEEEtran.bst: No hyphenation pattern has been}%
\typeout{** loaded for the language `#1'. Using the pattern for}%
\typeout{** the default language instead.}%
\else
\language=\csname l@#1\endcsname
\fi
#2}}
\providecommand{\BIBdecl}{\relax}
\BIBdecl

\bibitem{dla_survey}
S.~Pouyanfar, S.~Sadiq, Y.~Yan, H.~Tian, Y.~Tao, M.~P. Reyes, M.-L. Shyu, S.-C. Chen, and S.~S. Iyengar, ``A survey on deep learning: Algorithms, techniques, and applications,'' \emph{ACM computing surveys (CSUR)}, vol.~51, no.~5, pp. 1--36, 2018.

\bibitem{vaswani2017attention}
A.~Vaswani, N.~Shazeer, N.~Parmar, J.~Uszkoreit, L.~Jones, A.~N. Gomez, {\L}.~Kaiser, and I.~Polosukhin, ``Attention is all you need,'' \emph{Advances in neural information processing systems}, vol.~30, 2017.

\bibitem{llm_survey}
P.~Kumar, ``Large language models (llms): survey, technical frameworks, and future challenges,'' \emph{Artificial Intelligence Review}, vol.~57, no.~10, p. 260, 2024.

\bibitem{trainableHD}
J.~Kim, H.~Lee, M.~Imani, and Y.~Kim, ``Advancing hyperdimensional computing based on trainable encoding and adaptive training for efficient and accurate learning,'' \emph{ACM Transactions on Design Automation of Electronic Systems}, vol.~29, no.~5, pp. 1--25, 2024.

\bibitem{survey-hdc-stoch-framework}
M.~Heddes, I.~Nunes, T.~Givargis, A.~Nicolau, and A.~Veidenbaum, ``Hyperdimensional computing: a framework for stochastic computation and symbolic ai,'' \emph{Journal of Big Data}, vol.~11, no.~1, p. 145, 2024.

\bibitem{kanerva}
P.~Kanerva, ``Hyperdimensional computing: An introduction to computing in distributed representation with high-dimensional random vectors,'' \emph{Cognitive computation}, vol.~1, pp. 139--159, 2009.

\bibitem{survey-two-parter-part-I}
D.~Kleyko, D.~A. Rachkovskij, E.~Osipov, and A.~Rahimi, ``A survey on hyperdimensional computing aka vector symbolic architectures, part i: Models and data transformations,'' \emph{ACM Computing Surveys}, vol.~55, no.~6, pp. 1--40, 2022.

\bibitem{gpu-imani-cascade-hd}
Y.~Kim, J.~Kim, and M.~Imani, ``Cascadehd: Efficient many-class learning framework using hyperdimensional computing,'' in \emph{2021 58th ACM/IEEE Design Automation Conference (DAC)}.\hskip 1em plus 0.5em minus 0.4em\relax IEEE, 2021, pp. 775--780.

\bibitem{rosing-vision-hd}
\BIBentryALTinterwordspacing
F.~Asgarinejad, J.~Morris, T.~Rosing, and B.~Aksanli, ``Visionhd: Towards efficient and privacy-preserved hyperdimensional computing for image data,'' in \emph{Proceedings of the 29th ACM/IEEE International Symposium on Low Power Electronics and Design}, ser. ISLPED '24.\hskip 1em plus 0.5em minus 0.4em\relax New York, NY, USA: Association for Computing Machinery, 2024, p. 1–6. [Online]. Available: \url{https://doi.org/10.1145/3665314.3670852}
\BIBentrySTDinterwordspacing

\bibitem{trad-hdc-1}
A.~Rahimi, P.~Kanerva, L.~Benini, and J.~M. Rabaey, ``Efficient biosignal processing using hyperdimensional computing: Network templates for combined learning and classification of exg signals,'' \emph{Proceedings of the IEEE}, vol. 107, no.~1, pp. 123--143, 2018.

\bibitem{trad-hdc-2}
A.~Rahimi, P.~Kanerva, and J.~M. Rabaey, ``A robust and energy-efficient classifier using brain-inspired hyperdimensional computing,'' in \emph{Proceedings of the 2016 international symposium on low power electronics and design}, 2016, pp. 64--69.

\bibitem{trad-hdc-3-imani-voice-hd}
M.~Imani, D.~Kong, A.~Rahimi, and T.~Rosing, ``Voicehd: Hyperdimensional computing for efficient speech recognition,'' in \emph{2017 IEEE International Conference on Rebooting Computing (ICRC)}, 2017, pp. 1--8.

\bibitem{trad-hdc-4-imani-rosing-hdna}
M.~Imani, T.~Nassar, A.~Rahimi, and T.~Rosing, ``Hdna: Energy-efficient dna sequencing using hyperdimensional computing,'' in \emph{2018 IEEE EMBS International Conference on Biomedical \& Health Informatics (BHI)}.\hskip 1em plus 0.5em minus 0.4em\relax IEEE, 2018, pp. 271--274.

\bibitem{trad-hdc-5-imani-rosing-hierarchical-hd}
M.~Imani, C.~Huang, D.~Kong, and T.~Rosing, ``Hierarchical hyperdimensional computing for energy efficient classification,'' in \emph{Proceedings of the 55th Annual Design Automation Conference}, 2018, pp. 1--6.

\bibitem{adv-encoding-1-hv-design-eff-hdc}
T.~Basaklar, Y.~Tuncel, S.~Y. Narayana, S.~Gumussoy, and U.~Y. Ogras, ``Hypervector design for efficient hyperdimensional computing on edge devices,'' \emph{arXiv preprint arXiv:2103.06709}, 2021.

\bibitem{adv-encoding-2-encoding-binarized-img}
L.~Smets, W.~Van~Leekwijck, I.~J. Tsang, and S.~Latr{\'e}, ``An encoding framework for binarized images using hyperdimensional computing,'' \emph{Frontiers in big data}, vol.~7, p. 1371518, 2024.

\bibitem{adv-encoding-3-hardware-aware-static-opt}
P.~Yi and S.~Achour, ``Hardware-aware static optimization of hyperdimensional computations,'' \emph{Proceedings of the ACM on Programming Languages}, vol.~7, no. OOPSLA2, pp. 1--30, 2023.

\bibitem{adv-encoding-4-tiny-hd}
B.~Khaleghi, H.~Xu, J.~Morris, and T.~{\v{S}}. Rosing, ``tiny-hd: Ultra-efficient hyperdimensional computing engine for iot applications,'' in \emph{2021 Design, Automation \& Test in Europe Conference \& Exhibition (DATE)}.\hskip 1em plus 0.5em minus 0.4em\relax IEEE, 2021, pp. 408--413.

\bibitem{static-encoding-1-programmable-hdc}
S.~Datta, R.~A. Antonio, A.~R. Ison, and J.~M. Rabaey, ``A programmable hyper-dimensional processor architecture for human-centric iot,'' \emph{IEEE Journal on Emerging and Selected Topics in Circuits and Systems}, vol.~9, no.~3, pp. 439--452, 2019.

\bibitem{static-encoding-2-seizure-det}
L.~Ge and K.~K. Parhi, ``Seizure detection using power spectral density via hyperdimensional computing,'' in \emph{ICASSP 2021-2021 IEEE International Conference on Acoustics, Speech and Signal Processing (ICASSP)}.\hskip 1em plus 0.5em minus 0.4em\relax IEEE, 2021, pp. 7858--7862.

\bibitem{static-encoding-3-colal-learning-secure-hdc}
M.~Imani, Y.~Kim, S.~Riazi, J.~Messerly, P.~Liu, F.~Koushanfar, and T.~Rosing, ``A framework for collaborative learning in secure high-dimensional space,'' in \emph{2019 IEEE 12th International Conference on Cloud Computing (CLOUD)}.\hskip 1em plus 0.5em minus 0.4em\relax IEEE, 2019, pp. 435--446.

\bibitem{pim-dual-accl-cluster-pim}
M.~Imani, S.~Pampana, S.~Gupta, M.~Zhou, Y.~Kim, and T.~Rosing, ``Dual: Acceleration of clustering algorithms using digital-based processing in-memory,'' in \emph{2020 53rd Annual IEEE/ACM International Symposium on Microarchitecture (MICRO)}.\hskip 1em plus 0.5em minus 0.4em\relax IEEE, 2020, pp. 356--371.

\bibitem{gpu-xcel-hd-efficient-gpu}
J.~Kang, B.~Khaleghi, Y.~Kim, and T.~Rosing, ``Xcelhd: An efficient gpu-powered hyperdimensional computing with parallelized training,'' in \emph{2022 27th Asia and South Pacific Design Automation Conference (ASP-DAC)}.\hskip 1em plus 0.5em minus 0.4em\relax IEEE, 2022, pp. 220--225.

\bibitem{static-encoding-6-genie-hd}
Y.~Kim, M.~Imani, N.~Moshiri, and T.~Rosing, ``Geniehd: Efficient dna pattern matching accelerator using hyperdimensional computing,'' in \emph{2020 Design, Automation \& Test in Europe Conference \& Exhibition (DATE)}.\hskip 1em plus 0.5em minus 0.4em\relax IEEE, 2020, pp. 115--120.

\bibitem{static-encoding-7-efficient-human-activity-recognition-hdc-imani}
Y.~Kim, M.~Imani, and T.~S. Rosing, ``Efficient human activity recognition using hyperdimensional computing,'' in \emph{Proceedings of the 8th International Conference on the Internet of Things}, 2018, pp. 1--6.

\bibitem{static-encoding-8-distri-HD}
D.~Liang, J.~Shiomi, N.~Miura, and H.~Awano, ``Distrihd: a memory efficient distributed binary hyperdimensional computing architecture for image classification,'' in \emph{2022 27th Asia and South Pacific Design Automation Conference (ASP-DAC)}.\hskip 1em plus 0.5em minus 0.4em\relax IEEE, 2022, pp. 43--49.

\bibitem{static-encoding-9-hdc-framework-image-descriptors-cvpr}
P.~Neubert and S.~Schubert, ``Hyperdimensional computing as a framework for systematic aggregation of image descriptors,'' in \emph{Proceedings of the IEEE/CVF conference on computer vision and pattern recognition}, 2021, pp. 16\,938--16\,947.

\bibitem{hetero-algorithm-hardware-co-design}
Y.~Ni, Y.~Kim, T.~Rosing, and M.~Imani, ``Algorithm-hardware co-design for efficient brain-inspired hyperdimensional learning on edge,'' in \emph{2022 Design, Automation \& Test in Europe Conference \& Exhibition (DATE)}.\hskip 1em plus 0.5em minus 0.4em\relax IEEE, 2022, pp. 292--297.

\bibitem{static-encoding-11-imani-scalable-edge-hdc}
Z.~Zou, Y.~Kim, F.~Imani, H.~Alimohamadi, R.~Cammarota, and M.~Imani, ``Scalable edge-based hyperdimensional learning system with brain-like neural adaptation,'' in \emph{Proceedings of the International Conference for High Performance Computing, Networking, Storage and Analysis}, 2021, pp. 1--15.

\bibitem{trainable-hd-old-version}
J.~Kim, H.~Lee, M.~Imani, and Y.~Kim, ``Efficient hyperdimensional learning with trainable, quantizable, and holistic data representation,'' in \emph{2023 Design, Automation \& Test in Europe Conference \& Exhibition (DATE)}.\hskip 1em plus 0.5em minus 0.4em\relax IEEE, 2023, pp. 1--6.

\bibitem{multi-mani-hd-imani}
Z.~Zou, Y.~Kim, M.~H. Najafi, and M.~Imani, ``Manihd: Efficient hyper-dimensional learning using manifold trainable encoder,'' in \emph{2021 Design, Automation \& Test in Europe Conference \& Exhibition (DATE)}.\hskip 1em plus 0.5em minus 0.4em\relax IEEE, 2021, pp. 850--855.

\bibitem{survey-two-parter-part-II}
D.~Kleyko, D.~Rachkovskij, E.~Osipov, and A.~Rahimi, ``A survey on hyperdimensional computing aka vector symbolic architectures, part ii: Applications, cognitive models, and challenges,'' \emph{ACM Computing Surveys}, vol.~55, no.~9, pp. 1--52, 2023.

\bibitem{survey-hdc-edge-intel-progress}
C.-Y. Chang, Y.-C. Chuang, C.-T. Huang, and A.-Y. Wu, ``Recent progress and development of hyperdimensional computing (hdc) for edge intelligence,'' \emph{IEEE Journal on Emerging and Selected Topics in Circuits and Systems}, vol.~13, no.~1, pp. 119--136, 2023.

\bibitem{fpga-e3-hdc}
M.~S. Roodsari, J.~Krautter, V.~Meyers, and M.~Tahoori, ``E 3 hdc: Energy efficient encoding for hyper-dimensional computing on edge devices,'' in \emph{2024 34th International Conference on Field-Programmable Logic and Applications (FPL)}.\hskip 1em plus 0.5em minus 0.4em\relax IEEE, 2024, pp. 274--280.

\bibitem{fpga-edge-ai-acc}
J.-Y. Li and W.-C. Fang, ``An edge ai accelerator design based on hdc model for real-time eeg-based emotion recognition system with risc-v fpga platform,'' in \emph{2024 IEEE International Symposium on Circuits and Systems (ISCAS)}.\hskip 1em plus 0.5em minus 0.4em\relax IEEE, 2024, pp. 1--5.

\bibitem{fpga-general-hdc-edge}
M.~Asghari and S.~Le~Beux, ``A general purpose hyperdimensional computing accelerator for edge computing,'' in \emph{2024 22nd IEEE Interregional NEWCAS Conference (NEWCAS)}.\hskip 1em plus 0.5em minus 0.4em\relax IEEE, 2024, pp. 383--387.

\bibitem{fpga-hdc-hardware-acc-cvpr}
I.~Kandaswamy, S.~Farkya, Z.~Daniels, G.~van~der Wal, A.~Raghavan, Y.~Zhang, J.~Hu, M.~Lomnitz, M.~Isnardi, D.~Zhang \emph{et~al.}, ``Real-time hyper-dimensional reconfiguration at the edge using hardware accelerators,'' in \emph{Proceedings of the IEEE/CVF Conference on Computer Vision and Pattern Recognition}, 2022, pp. 3610--3618.

\bibitem{fpga-imani-revisiting-hdc-fpga}
M.~Imani, Z.~Zou, S.~Bosch, S.~A. Rao, S.~Salamat, V.~Kumar, Y.~Kim, and T.~Rosing, ``Revisiting hyperdimensional learning for fpga and low-power architectures,'' in \emph{2021 IEEE International Symposium on High-Performance Computer Architecture (HPCA)}.\hskip 1em plus 0.5em minus 0.4em\relax IEEE, 2021, pp. 221--234.

\bibitem{fpga-onlinehd}
A.~Hern{\'a}ndez-Cano, N.~Matsumoto, E.~Ping, and M.~Imani, ``Onlinehd: Robust, efficient, and single-pass online learning using hyperdimensional system,'' in \emph{2021 Design, Automation \& Test in Europe Conference \& Exhibition (DATE)}.\hskip 1em plus 0.5em minus 0.4em\relax IEEE, 2021, pp. 56--61.

\bibitem{fpga-scalable-interpretable}
H.~Chen, Y.~Ni, W.~Huang, and M.~Imani, ``Scalable and interpretable brain-inspired hyper-dimensional computing intelligence with hardware-software co-design,'' in \emph{2024 IEEE Custom Integrated Circuits Conference (CICC)}.\hskip 1em plus 0.5em minus 0.4em\relax IEEE, 2024, pp. 1--8.

\bibitem{gpu-hdtorch}
\BIBentryALTinterwordspacing
W.~A. Simon, U.~Pale, T.~Teijeiro, and D.~Atienza, ``Hdtorch: Accelerating hyperdimensional computing with gp-gpus for design space exploration,'' in \emph{Proceedings of the 41st IEEE/ACM International Conference on Computer-Aided Design}, ser. ICCAD ’22.\hskip 1em plus 0.5em minus 0.4em\relax ACM, Oct. 2022, p. 1–8. [Online]. Available: \url{http://dx.doi.org/10.1145/3508352.3549475}
\BIBentrySTDinterwordspacing

\bibitem{gpu-openhd}
J.~Kang, B.~Khaleghi, T.~Rosing, and Y.~Kim, ``Openhd: A gpu-powered framework for hyperdimensional computing,'' \emph{IEEE Transactions on Computers}, vol.~71, no.~11, pp. 2753--2765, 2022.

\bibitem{pim-biohd-imani}
Z.~Zou, H.~Chen, P.~Poduval, Y.~Kim, M.~Imani, E.~Sadredini, R.~Cammarota, and M.~Imani, ``Biohd: an efficient genome sequence search platform using hyperdimensional memorization,'' in \emph{Proceedings of the 49th Annual International Symposium on Computer Architecture}, 2022.

\bibitem{pim-hdnn-rosing}
A.~Dutta, S.~Gupta, B.~Khaleghi, R.~Chandrasekaran, W.~Xu, and T.~Rosing, ``Hdnn-pim: Efficient in memory design of hyperdimensional computing with feature extraction,'' in \emph{Proceedings of the Great Lakes Symposium on VLSI 2022}, 2022, pp. 281--286.

\bibitem{pim-paap-hd}
F.~Liu, H.~Li, N.~Yang, Y.~Chen, Z.~Wang, T.~Yang, and L.~Jiang, ``Paap-hd: Pim-assisted approximation for efficient hyper-dimensional computing,'' in \emph{2024 29th Asia and South Pacific Design Automation Conference (ASP-DAC)}.\hskip 1em plus 0.5em minus 0.4em\relax IEEE, 2024, pp. 46--51.

\bibitem{hetero-hpvm-hdc}
R.~Arbore, X.~Routh, A.~R. Noor, A.~Kothari, H.~Yang, W.~Xu, S.~Pinge, V.~Adve, T.~Rosing, and M.~Zhou, ``Hpvm-hdc: A heterogeneous programming system for accelerating hyperdimensional computing,'' \emph{arXiv preprint arXiv:2410.15179}, 2024.

\bibitem{multi-comphd}
J.~Morris, M.~Imani, S.~Bosch, A.~Thomas, H.~Shu, and T.~Rosing, ``Comphd: Efficient hyperdimensional computing using model compression,'' in \emph{2019 IEEE/ACM International Symposium on Low Power Electronics and Design (ISLPED)}.\hskip 1em plus 0.5em minus 0.4em\relax IEEE, 2019, pp. 1--6.

\bibitem{multi-multiarch-hardware-acc}
I.~Peitzsch, M.~Ciora, and A.~D. George, ``Multiarchitecture hardware acceleration of hyperdimensional computing,'' in \emph{2023 IEEE High Performance Extreme Computing Conference (HPEC)}.\hskip 1em plus 0.5em minus 0.4em\relax IEEE, 2023, pp. 1--7.

\bibitem{cpu-hdcc-compiler}
P.~Verg{\'e}s, M.~Heddes, I.~Nunes, T.~Givargis, and A.~Nicolau, ``Hdcc: A hyperdimensional computing compiler for classification on embedded systems and high-performance computing,'' \emph{arXiv preprint arXiv:2304.12398}, 2023.

\bibitem{advance-on-cpu-ref}
E.~Hassan, M.~Bettayeb, and B.~Mohammad, ``Advancing hardware implementation of hyperdimensional computing for edge intelligence,'' in \emph{2024 IEEE 6th International Conference on AI Circuits and Systems (AICAS)}.\hskip 1em plus 0.5em minus 0.4em\relax IEEE, 2024, pp. 169--173.

\bibitem{graph-hdgl}
\BIBentryALTinterwordspacing
A.~Dalvi and V.~Honavar, ``Hyperdimensional representation learning for node classification and link prediction,'' in \emph{Proceedings of the Eighteenth ACM International Conference on Web Search and Data Mining}, ser. WSDM ’25.\hskip 1em plus 0.5em minus 0.4em\relax ACM, Mar. 2025, p. 88–97. [Online]. Available: \url{http://dx.doi.org/10.1145/3701551.3703492}
\BIBentrySTDinterwordspacing

\bibitem{adam}
\BIBentryALTinterwordspacing
D.~P. Kingma and J.~Ba, ``Adam: A method for stochastic optimization,'' 2017. [Online]. Available: \url{https://arxiv.org/abs/1412.6980}
\BIBentrySTDinterwordspacing

\bibitem{qat-1}
\BIBentryALTinterwordspacing
S.~K. Esser, J.~L. McKinstry, D.~Bablani, R.~Appuswamy, and D.~S. Modha, ``Learned step size quantization,'' 2020. [Online]. Available: \url{https://arxiv.org/abs/1902.08153}
\BIBentrySTDinterwordspacing

\bibitem{qat-2}
\BIBentryALTinterwordspacing
M.~Nagel, M.~Fournarakis, R.~A. Amjad, Y.~Bondarenko, M.~van Baalen, and T.~Blankevoort, ``A white paper on neural network quantization,'' 2021. [Online]. Available: \url{https://arxiv.org/abs/2106.08295}
\BIBentrySTDinterwordspacing

\bibitem{intel-simd}
I.~Corporation, ``Improve performance with intel avx-512 vectorization,'' \url{https://www.intel.com/content/www/us/en/developer/articles/technical/improve-performance-with-vectorization.html}, 2023, accessed: 2024-05-13.

\bibitem{emotion-det}
J.~J. Bird, A.~Ekart, C.~D. Buckingham, and D.~R. Faria, ``Mental emotional sentiment classification with an eeg-based brain-machine interface,'' in \emph{Proceedings of theInternational Conference on Digital Image and Signal Processing (DISP’19)}, 2019.

\bibitem{activity-rec-1}
A.~Reiss and D.~Stricker, ``Introducing a new benchmarked dataset for activity monitoring,'' in \emph{2012 16th international symposium on wearable computers}.\hskip 1em plus 0.5em minus 0.4em\relax IEEE, 2012, pp. 108--109.

\bibitem{activity-rec-2}
J.-L. Reyes-Ortiz, L.~Oneto, A.~Sam{\`a}, X.~Parra, and D.~Anguita, ``Transition-aware human activity recognition using smartphones,'' \emph{Neurocomputing}, vol. 171, pp. 754--767, 2016.

\bibitem{class-3}
Y.~LeCun, L.~Bottou, Y.~Bengio, and P.~Haffner, ``Gradient-based learning applied to document recognition,'' \emph{Proceedings of the IEEE}, vol.~86, no.~11, pp. 2278--2324, 1998.

\bibitem{class-4}
C.~Mallah, J.~Cope, J.~Orwell \emph{et~al.}, ``Plant leaf classification using probabilistic integration of shape, texture and margin features,'' \emph{Signal Processing, Pattern Recognition and Applications}, vol.~5, no.~1, pp. 45--54, 2013.

\bibitem{hact}
\BIBentryALTinterwordspacing
D.~Dua and C.~Graff, ``Uci machine learning repository,'' 2017. [Online]. Available: \url{http://archive.ics.uci.edu/ml}
\BIBentrySTDinterwordspacing

\bibitem{ecg-heart}
M.~Kachuee, S.~Fazeli, and M.~Sarrafzadeh, ``Ecg heartbeat classification: A deep transferable representation,'' in \emph{2018 IEEE international conference on healthcare informatics (ICHI)}.\hskip 1em plus 0.5em minus 0.4em\relax IEEE, 2018, pp. 443--444.

\bibitem{pthreads}
{IEEE Std 1003.1}, ``Posix threads programming,'' 2004, \url{https://pubs.opengroup.org/onlinepubs/9699919799/functions/pthread_create.html}.

\bibitem{openmp}
{OpenMP Architecture Review Board}, \emph{OpenMP Application Programming Interface Version 5.0}, \url{https://www.openmp.org/specifications/}, OpenMP, 2018, accessed: 2025-06-03.

\bibitem{concurrentqueue}
cameron314, ``Concurrentqueue: A fast multi-producer, multi-consumer lock-free queue for c++,'' \url{https://github.com/cameron314/concurrentqueue}, 2016, accessed: 2025-06-03.

\bibitem{eigen}
G.~Guennebaud, B.~Jacob \emph{et~al.}, ``Eigen: C++ template library for linear algebra,'' \url{https://eigen.tuxfamily.org}, 2023, version 3.4.0, Accessed: 2025-06-03.

\bibitem{torchhd}
\BIBentryALTinterwordspacing
M.~Heddes, I.~Nunes, P.~Vergés, D.~Kleyko, D.~Abraham, T.~Givargis, A.~Nicolau, and A.~Veidenbaum, ``Torchhd: An open source python library to support research on hyperdimensional computing and vector symbolic architectures,'' \emph{Journal of Machine Learning Research}, vol.~24, no. 255, pp. 1--10, 2023. [Online]. Available: \url{http://jmlr.org/papers/v24/23-0300.html}
\BIBentrySTDinterwordspacing

\bibitem{pytorch}
{PyTorch Core Team}, ``Pytorch,'' 2024, \url{https://pytorch.org/}.

\bibitem{imani-brain-HDC}
H.~Amrouch, M.~Imani, X.~Jiao, Y.~Aloimonos, C.~Fermuller, D.~Yuan, D.~Ma, H.~E. Barkam, P.~R. Genssler, and P.~Sutor, ``Brain-inspired hyperdimensional computing for ultra-efficient edge ai,'' in \emph{2022 International Conference on Hardware/Software Codesign and System Synthesis (CODES+ ISSS)}.\hskip 1em plus 0.5em minus 0.4em\relax IEEE, 2022, pp. 25--34.

\end{thebibliography}

\end{document}